\documentclass[preprint,12pt]{elsarticle}
\usepackage{amssymb}
\usepackage{amsmath}
\usepackage[nolist]{acronym}
\usepackage[hyphens]{url}
\usepackage[hidelinks]{hyperref}
\hypersetup{breaklinks=true}
\journal{Computer Physics Communications}

\newtheorem{hypothesis}{Hypothesis}
\newtheorem{definition}{Definition}
\newtheorem{lemma}{Lemma}
\newtheorem{corollary}{Corollary}

\acrodef{PUMAS}{Semi-Analytical MUons Propagation}
\acrodef{CSDA}{Continuously Slowing Down Approximation}
\acrodef{TT}{Transverse Transport}
\acrodef{DCS}{Differential Cross-Section}
\acrodef{PDG}{Particle Data Group}
\acrodef{pdf}{probability density function}
\acrodef{cdf}{cumulative distribution function}
\acrodef{MC}{Monte-Carlo}
\acrodef{AMC}{Adjoint Monte-Carlo}
\acrodef{BMC}{Backward Monte-Carlo}
\acrodef{DEL}{Discrete Energy Loss}
\acrodef{CEL}{Continuous Energy Loss}

\begin{document}

\begin{frontmatter}
\title{Backward Monte-Carlo applied to muon transport}

\author[lpc]{V.~Niess\corref{cor1}}
\ead{niess@in2p3.fr}
\author[lpc]{A.~Barnoud}
\author[lpc]{C.~C\^arloganu}
\author[lpc]{E.~Le~Menedeu}

\cortext[cor1]{Corresponding author}

\address[lpc]{Universit\'e~Clermont Auvergne, CNRS/IN2P3, LPC, F-63000 Clermont-Ferrand, France.}

\begin{abstract}
We discuss a backward Monte-Carlo technique for muon transport problem, with emphasis on its application in muography. Backward Monte-Carlo allows exclusive sampling of a final state by reversing the simulation flow. In practice it can be made analogous to an adjoint Monte-Carlo, though it is more versatile for muon transport. A backward Monte-Carlo was implemented as a dedicated muon transport library: PUMAS. It is shown for case studies relevant for muography imaging that the implementations of forward and backward Monte-Carlo schemes agree to  better than $1$~\%.
\end{abstract}

\begin{keyword}
adjoint Monte-Carlo \sep backward Monte-Carlo \sep muon transport.



\end{keyword}

\end{frontmatter}


\section{Introduction}
In the present section we start by giving the motivation for this work and by setting the context. Section~\ref{sec:backward_mc} is dedicated to a general discussion of the \ac{BMC} technique as applied to the muon transport problem. Then, a specific implementation is detailed in section~\ref{sec:implementation}. Finally, validation results can be found in section~\ref{sec:validation}.

\subsection{Motivation}
This work has been largely motivated by the muography imaging. Muography is a radiography technique for large scale, $0.1-2$~km, dense structures, e.g. pyramids~\cite{Alvarez:1975, Menchaca-Rocha:2014}, small volcanoes~\cite{Nagamine:1995, Lesparre:2012, Portal:2013, Ambrosino:2014, Tanaka:2014}, {\it etc} .... The measurement of the transmitted flux of atmospheric muons through a target allows inferring the integrated density along lines of sight. A precise density measurement requires dedicated \ac{MC} computations. These computations can be very inefficient since in a classical \ac{MC} one has to sample muons produced over the whole atmosphere but going through a tiny $\simeq 1 \text{m}^2$ detection plane. This is especially problematic for low energy muons that can scatter away a lot from their initial directions at production. Although those non ballistic muons are not of direct interest for muography they contribute -as well as other particles- to the background~\cite{Ambrosino:2015, Nishiyama:2016}. Therefore it is desirable to account for them by simulation, if they cannot be rejected experimentally. A successful strategy in such cases is to reverse the simulation flow using \ac{AMC} techniques. This guarantees to sample only useful events by starting from the detector and going backwards to the source.

So far there has not been any implementation of \ac{AMC} for muography. The closest work is the one of Desorgher {\it et al.}~\citep{Desorgher:2010} with an implementation in Geant4~\cite{Geant4} for space radiation problems, i.e. $e^{\pm}$, $p$ and $\gamma$. It would have been possible to extend the work of Desorgher {\it et al.} to muons. But a dedicated code that would allow specific optimisations for muography was thought to be more efficient. Following this path triggered an alternative scheme for the backward sampling of muons, which is detailed herein.

The starting idea of the \ac{BMC} is to directly reverse the simulation algorithmic step by step. For example, given a random procedure to generate a final state from an initial state and some random variates, one simply reverses it to express the initial state as function of the final state. When the inverse exists this is a correct procedure up to a weight factor for the conservation of the probability density, as detailed in lemma~\ref{lm:backward_sampling} of section~\ref{sec:backward_mc}. Applying this method to the muon transport problem one gets an algorithm very close to the \ac{AMC} of Desorger {\it et al.} though more versatile. Since it does not rely on an adjoint formulation we call this method backward Monte-Carlo. The term reverse Monte-Carlo also found in the literature would have been misleading to us because this technique does not solve any inverse problem. Indeed, backward sampling a bunch of scattered particles will not re-focus them on the initial source. Instead, \ac{BMC} deals with the exact same problem as the forward case but it allows to generate only useful events, i.e. final states that are actually observed.

\subsection{Definitions}
For the present discussion a \ac{MC} state is simplified as a set $\mathbf{s} = \{\epsilon, \vec{r}, \vec{u}\}$ describing the components of a stationary flux of relativistic muons of kinetic energy $\epsilon$, at the space coordinate $\vec{r}$ and with momentum's direction $\vec{u}$. The muons interactions with matter are split into a continuous component describing collective processes, e.g. multiple scattering and \ac{CEL}, and one or more discrete processes leading to catastrophic interactions. The following is assumed throughout this work:
\begin{hypothesis}[H\ref{hyp:base}] \ \
\label{hyp:base}
\begin{enumerate}[(i)]
{\item the work of all external forces has a null sum at every space coordinate.}
{\item the medium density is constant or atomic cross-sections do not depend on the medium density.}
{\item the differential cross-sections are invariant by rotation, e.g. unpolarised.}
\end{enumerate}
\end{hypothesis}
In practice, atomic cross-sections do actually depend on the medium density, due for example to the density effect in ionisation loss. However, in some circumstances this might be negligible, e.g. when considering a porous material with minerals of constant density and pores of air. Then, the bulk density variations actually reflect porosity variations while the energy loss is dominated by the minerals of constant density.

Note that in a realistic case the simulated geometry actually consists of several media with different compositions and densities, e.g. rocks and air in muography. However, let us first consider a single medium and then generalise to a succession of different media, provided that they all satisfy to hypothesis~\ref{hyp:base}.

Under hypothesis~\ref{hyp:base}, the continuous energy loss depends only on the column density along the path, $x = \int{\rho ds}$. Therefore, the \ac{CEL} could be generated {\it a priori}, parametrised only by the column density, although this is not a realistic \ac{MC} implementation. Thus, it is conceptually convenient to model it as following:
\begin{definition}[Continuous energy loss]
\label{def:CEL}
A random function $\mathcal{E}(x)$ of the column density $x$ generates a continuous energy loss from $\epsilon_0$ to $\epsilon_1$,  with $\epsilon_0 > \epsilon_1 \geq 0$, if and only if there exists $x_1 > 0$ such that :
\begin{enumerate}[(i)]
{\item $\mathcal{E}(0) = \epsilon_0$ and $\mathcal{E}(x_1) = \epsilon_1$.}
{\item $\mathcal{E}$ is a strictly decreasing function over $\mathbb{I} = [0, x_1]$.}
{\item $\mathcal{E}$ is continuous and differentiable over $\mathbb{I}$.}
\end{enumerate}
Following, $\mathcal{E}^{-1}(\epsilon)$ the inverse of $\mathcal{E}$ over $\mathbb{I}$ exists. The continuous energy loss random function, $\mathcal{B}$, is defined over $[\epsilon_1, \epsilon_0]$ as:
\begin{equation}
\mathcal{B}(\epsilon) = -\frac{\partial \mathcal{E}}{\partial x}(\mathcal{E}^{-1}(\epsilon))
\end{equation}
where it must be understood that $\mathcal{E}$, $\partial \mathcal{E} / \partial x$ and $\mathcal{E}^{-1}$ refer to the same random variate.
\end{definition}
Let us point out that we follow the usual convention that a random variable or a random function is written with a capital letter while a given outcome is written with the corresponding lower case letter.

The previous definition for \ac{CEL} is purely mathematical. It states that \ac{CEL} is indeed sustained by a continuous variation of the energy along the particle trajectory and that there is a bijective mapping between kinetic energy and column density variations. It is up to the implementation to ensure that the corresponding \ac{CEL} actually observes the Physics. Given definition~\ref{def:CEL}, one can associate to any \ac{MC} particle a unique \ac{CEL} occurrence, $\beta(\epsilon)$, fully pre-generated at the start according to $\mathcal{B}$. Again this is not a realistic \ac{MC} procedure to implement but it is conceptually convenient and asymptotically exact. A realistic \ac{MC} procedure would actually compute the \ac{CEL} on the flight and might implement discontinuous effects during its stepping procedure as an approximation.

For discrete processes, on the contrary the kinetic energy and direction vary non continuously at some space coordinates, i.e. at vertices where interactions occur. In the following, let us call \ac{DEL} these processes. Let us also write $\lambda > 0$ the total interaction length for discrete processes. Under hypothesis~\ref{hyp:base}, the interaction length depends only on the kinetic energy. As a result it is noteworthy that the survival probability to \ac{DEL} does not depend on the geometric details of the \ac{CEL}, but only on the kinetic energy, as:
\begin{equation}
\label{eq:survival_probability}
P_{\lambda,\mathcal{B}}(\epsilon_f; \epsilon_i) = \exp(-\int_{\epsilon_f}^{\epsilon_i}{\frac{d\epsilon}{\lambda(\epsilon) \mathcal{B}(\epsilon)}})
\end{equation}
Note that $P_{\lambda,\mathcal{B}}$ is a random function again, derived from $\mathcal{B}$. This result is well known in the case of a deterministic \ac{CEL}, e.g. see Appendix~B of Sokalski {\it et al.}~\cite{MuM01}. It is also true for a stochastic \ac{CEL} provided that it satisfies definition~\ref{def:CEL}.

In addition, in order to simplify the present discussion it is convenient to make the following assumption:
\begin{hypothesis}[H\ref{hyp:direct_inverse}]
\label{hyp:direct_inverse}
The total macroscopic cross-section $\sigma = 1/\lambda$ is strictly positive for all values of the kinetic energy in $\mathbb{R}_+ = [0, +\infty[$.
\end{hypothesis}
Note that hypothesis~\ref{hyp:direct_inverse} is not rigorously true in a practical implementation, as for muon transport. However, a simple way to patch this is to regularise the total cross-section by adding a {\it do nothing} sub-process, as illustrated in section~\ref{sec:implementation}. But, this only shifts the problem to the sub-process sampling. Under hypothesis~\ref{hyp:direct_inverse}, $P_{\lambda,\mathcal{B}}$ can be inverted over $\mathbb{R}_+$ with respect to either $\epsilon_i$ or $\epsilon_f$. Consequently the forward and backward sampling procedures are symmetrical.

\section{Backward Monte-Carlo}
\label{sec:backward_mc}
The following is a general discussion of the \ac{BMC} technique. It provides a collection of lemmas allowing to build an end to end backward Monte-Carlo simulation of high energy muons. In order not to dilute the main results all proofs have been relegated to the appendices.

\subsection{Formulation}
The key element of the \ac{BMC} method is given by lemma~\ref{lm:backward_sampling}. It states that if one can invert the \ac{MC} process for generating the final state, $\mathbf{s}_f$, from the initial state, $\mathbf{s}_i$, the inverse process provides an exclusive \ac{MC} estimate of the final state density, after applying a Jacobian weight factor $|\partial \mathbf{s}_i / \partial \mathbf{s}_f|$.

\begin{lemma}
\label{lm:backward_sampling}
Let $\tau_{i,f}(\mathbf{s}_f; \mathbf{s}_i)$ be the probability density function, with respect to $\mathbf{s}_f$, of a transition probability from an initial state $\mathbf{s}_i$ to a final state $\mathbf{s}_f$. Let $\mathbf{s}_f = g(\mathbf{s}_i; X)$ be a Monte-Carlo process for generating $\mathbf{s}_f$ distributed as $\tau_{i,f}$, given $\mathbf{s}_i$ and a set of random variables $X$ that do not depend on $\mathbf{s}_i$. If $g^{-1}$ the inverse of $g$ with respect to $\mathbf{s}_i$ exists, then for any probability density function $\rho_i$ for the initial state, a backward Monte-Carlo estimate of the probability density function $\rho_f$ of the final state is:
\begin{equation}
\begin{split}
\rho_f(\mathbf{s}_f) = & \int{\tau_{i,f}(\mathbf{s}_f; \mathbf{s}_i) \rho_i(\mathbf{s}_i) d\mathbf{s}_i} \simeq \frac{1}{N} \sum_{k=1}^N{\omega_{i,k} \rho_i(\mathbf{s}_{i,k})} \\
\mathbf{s}_{i, k} = & g^{-1}(\mathbf{s}_f; x_k) \\
\omega_{i,k} = & \det(J_{g^{-1}, \mathbf{s}_f}(\mathbf{s}_f; x_k))
\end{split}
\end{equation}
where $x_k$ is a random variate drawn from $X$ and with $J_{g^{-1}, \mathbf{s}_f}$, informally $\partial \mathbf{s}_i / \partial \mathbf{s}_f$, the Jacobian matrix of $g^{-1}$ with respect to $\mathbf{s}_f$.
\end{lemma}
Note that lemma~\ref{lm:backward_sampling} is still valid when the state transition is a deterministic process, i.e. $X=\varnothing$. In this case the initial states $\mathbf{s}_{i, k}$ and their weights $\omega_{i, k}$ are all equal and one simply recovers the classical conservation law for a change of variable in a \ac{pdf}, i.e. $\rho_f \mathrm{d}\mathbf{s}_f = \rho_i \mathrm{d}\mathbf{s}_i$. The non obvious result stated here is that the same can be done event by event treating the underlying random variates, $x_k$, as constants. The proof is given in~\ref{sec:proof_1}.

Let us illustrate the method with a simple case useful for muon transport. Let us consider a unidimensional propagation with a deterministic \ac{CEL}, $\beta$, e.g. as implemented in MUM~\cite{MuM01}. Considering a muon with initial kinetic energy $\epsilon_i$, under hypothesis~\ref{hyp:direct_inverse}, the final kinetic energy $\epsilon_f$ at which the next stochastic interaction would occur can be randomised as:
\begin{equation}
\label{eq:forward_MUM}
\epsilon_f = P_{\lambda, \beta}^{-1}(P_{\lambda, \beta}(\epsilon_i)+\ln(x))
\end{equation}
with $x$ a random variate drawn from a uniform distribution over $[0,1]$ and where $P_{\lambda, \beta}(\epsilon) \equiv P_{\lambda, \beta}(\epsilon;0)$ was given in equation~(\ref{eq:survival_probability}). Inverting equation~(\ref{eq:forward_MUM}) for $\epsilon_i$ is straightforward, yielding:
\begin{equation}
\label{eq:interaction_backward_kinetic}
\epsilon_i = P_{\lambda, \beta}^{-1}(P_{\lambda, \beta}(\epsilon_f)-\ln(x))
\end{equation}
from which the Jacobian weight factor can be computed directly as:
\begin{equation}
\label{eq:interaction_backward_jacobian}
\left| \frac{\partial \epsilon_i}{\partial \epsilon_f} \right| = \frac{\lambda(\epsilon_i)\beta(\epsilon_i)}{\lambda(\epsilon_f)\beta(\epsilon_f)}
\end{equation}
Hence, in order to sample the final density for a given kinetic energy $\epsilon_f$ one simply needs to draw initial values $\epsilon_i$ from equation~(\ref{eq:interaction_backward_kinetic}) and to average the corresponding initial densities, $\rho_i$, applying the weight factor of equation~(\ref{eq:interaction_backward_jacobian}). Note that the forward \ac{MC} algorithm would instead sample $\epsilon_i$ from $\rho_i$ and then average the corresponding transition probabilities, $\tau_{i,f}$.

Given only the previous example, there would not be a significant difference between the forward and backward sampling. However, backward sampling becomes of interest when considering a succession of $n$ transition processes from state $\mathbf{s}_0$ to state $\mathbf{s}_n$. Let us write $\tau_{j,j+1}$ the density probability for a transition from $\mathbf{s}_j$ to $\mathbf{s}_{j+1}$ and $\mathbf{s}_{j+1} = g_j(\mathbf{s}_j; X_j)$ the corresponding \ac{MC} process. Then, applying lemma~\ref{lm:backward_sampling} recursively, the \ac{BMC} estimate of the final state density $\rho_n(\mathbf{s}_n)$ is given as:
\begin{equation}
\label{eq:backward_monte_carlo}
\begin{split}
\rho_n(\mathbf{s}_n) = & \int \tau_{n-1,n} d\mathbf{s}_{n-1} \int \tau_{n-2,n-1} d\mathbf{s}_{n-2}\ ...\ \int \tau_{0,1} \rho_0(\mathbf{s}_0) d\mathbf{s}_0 \\
\simeq & \frac{1}{N} \sum_j{\rho_0(\mathbf{s}_{0,j}) w_{0,j}} \\
s_{0,j} = & g_0^{-1} \circ\ ...\ g^{-1}_{n-2} \circ g^{-1}_{n-1}(\mathbf{s}_n; x_{n-1, j}) \\
w_{0,j} = & \prod_{k=1}^n{\det(J_{g^{-1}_{k-1}, \mathbf{s}_k})} = \left| \frac{\partial \mathbf{s}_{0,j}}{\partial \mathbf{s}_n} \right|
\end{split}
\end{equation}
with $\rho_0$ the initial density. Equation~(\ref{eq:backward_monte_carlo}) states that in order to estimate the final state density for a given state $\mathbf{s}_n$ one applies, in reverse order, the inverse \ac{MC} processes at each step as well as the corresponding Jacobian weight factors. By doing so, one gets a \ac{MC} initial state that is used to sample the initial state density. In comparison, if only a given final state $\mathbf{s}_n$ is of interest, a forward \ac{MC} estimate can be very inefficient since $\mathbf{s}_{n-1}$ might deviate significantly from the desired final state $\mathbf{s}_n$ after $n-1$ steps.

\subsection{Biasing}
In practice \ac{MC} processes can be complicated to invert. Hopefully, as for the forward \ac{MC}, one can substitute the true transition distribution with an approximate but more compliant one. This leads to the following biasing variant of lemma~\ref{lm:backward_sampling}.
\begin{corollary}
\label{cor:backward_prior_sampling}
Let $\tau_{i,f}^{(b)}(\mathbf{s}_f; \mathbf{s}_i)$ be a probability density function for $\mathbf{s}_f$ such that $g_b^{-1}$, the inverse of $\mathbf{s}_f = g_b(\mathbf{s}_i; X)$ with respect to $\mathbf{s}_i$, exists. Then, a Monte-Carlo estimate of $\rho_f(\mathbf{s}_f)$ is:
\begin{equation}
\begin{split}
\rho_f(\mathbf{s}_f) = & \int{ \tau_{i,f}(\mathbf{s}_f; \mathbf{s}_i) \rho_i(\mathbf{s}_i) d\mathbf{s}_i} \simeq \frac{1}{N} \sum_{k=1}^N{\rho_i(\mathbf{s}_{i,k}) \omega_{i,k}} \\
\mathbf{s}_{i, k} = & g_b^{-1}(\mathbf{s}_f; x_k) \\
\omega_{i,k} = &  \frac{\tau_{i,f}(\mathbf{s}_f; \mathbf{s}_{i,k})}{\tau_{i,f}^{(b)}(\mathbf{s}_f; \mathbf{s}_{i,k})} J_{g_b^{-1}, \mathbf{s}_f}(\mathbf{s}_f; x_k)
\end{split}
\end{equation}
provided that $\tau_{i,f}^{(b)} > 0$ for any $\mathbf{s}_i$ where $\tau_{i,f} > 0$.
\end{corollary}
The proof is almost identical to the previous proof of lemma~\ref{lm:backward_sampling}, hence it is not detailed in this paper. Corollary~\ref{cor:backward_prior_sampling} allows to postulate an {\it a priori} distribution for the transition probability, easier to invert, but correcting {\it a posteriori} for the true distribution value. Let us point out that the weight factor conveniently involves the Jacobian of the bias \ac{MC} process, $g_b$, not the true one, $g$. Thus, combined with biasing, \ac{BMC} becomes simple to implement.

A case of particular interest is a mixture \ac{pdf} where a same final state could be reached through several components. Let us write such a \ac{pdf} as:
\begin{equation}
\label{eq:mixture_pdf}
\tau_{i,f}(\mathbf{s}_f; \mathbf{s}_i) = \sum_{\ell=1}^m{p_\ell(\mathbf{s}_i) \tau_{i,f}^{(\ell)}(\mathbf{s}_f; \mathbf{s}_i)}
\end{equation}
with $p_\ell \in [0, 1]$ the weights of the mixture and $\sum_\ell p_\ell = 1$. In a forward sampling one would first draw out a component, let's say $\ell_0$, according to the probabilities $p_\ell$. Then, one generates the final state from $\mathbf{s}_f = g_{\ell_0}(\mathbf{s}_i; X_{\ell_0})$. When the mixture weights depend on the initial state this procedure cannot be inverted directly. A first solution is to use an approximation to the mixture, but invertible, and rely on corollary~\ref{cor:backward_prior_sampling} to draw the initial state. Then one can proceed as in the forward case and draw the mixture component. This can be rough though in some cases. Instead, in a backward sampling one might have a good {\it a priori} estimate of the mixture weights even though the initial state is not known. If so, corollary~\ref{cor:backward_mixture_sampling} hereafter provides an efficient bias method for the backward sampling of the mixture. Note that one might use the exact mixture components if they are easily inverted. Again the proof is similar to the one of lemma~\ref{lm:backward_sampling}.
\begin{corollary}
\label{cor:backward_mixture_sampling}
Let $\tau_{i,f}(\mathbf{s}_f; \mathbf{s}_i)$ be a mixture \ac{pdf} for $\mathbf{s}_f$ given by equation~(\ref{eq:mixture_pdf}). Let us write $\tau_{i,f}^{(\ell, b)}(\mathbf{s}_f; \mathbf{s}_i)$ a set of $m$ \ac{pdf} for $\mathbf{s}_f$ such that $g_{\ell, b}^{-1}$, the inverse of $\mathbf{s}_f = g_{\ell, b}(\mathbf{s}_i; X_\ell)$ with respect to $\mathbf{s}_i$, exists. In addition, let us write $p_{\ell, b}(\mathbf{s}_f)$ an estimate of the probability to select the $\ell^\text{th}$ component in a backward sampling. Then, provided that $p_{\ell, b} \tau_{i,f}^{(\ell, b)} > 0$ for any $\mathbf{s}_i$ where $p_\ell \tau^{(\ell)}_{i,f} > 0$, a Monte-Carlo estimate of $\rho_f(\mathbf{s}_f)$ is:
\begin{equation}
\begin{split}
\rho_f(\mathbf{s}_f) = & \int{ \tau_{i,f}(\mathbf{s}_f; \mathbf{s}_i) \rho_i(\mathbf{s}_i) d\mathbf{s}_i} \simeq \frac{1}{N} \sum_{k=1}^N{\rho_i(\mathbf{s}_{i,k}) \omega_{i,k}} \\
\mathbf{s}_{i, k} = & g_{\ell_0, b}^{-1}(\mathbf{s}_f; x_{\ell_0, k}) \\
\omega_{i,k} = &  \frac{p_{\ell_0}(\mathbf{s}_{i,k}) \tau^{(\ell_0)}_{i,f}(\mathbf{s}_f; \mathbf{s}_{i,k})}{p_{\ell_0, b}(\mathbf{s}_f) \tau_{i,f}^{(\ell_0, b)}(\mathbf{s}_f; \mathbf{s}_{i,k})} J_{g_{\ell_0, b}^{-1}, \mathbf{s}_f}(\mathbf{s}_f; x_{\ell_0, k})
\end{split}
\end{equation}
where the index $\ell_0(k)$ is drawn according to the probabilities $p_{\ell, b}(\mathbf{s}_f)$ and where $x_{\ell_0, k}$ is a random variate drawn from $X_{\ell_0}$.
\end{corollary}

\subsection{Transport}
Let us consider more in detail the inversion of the \ac{MC} process for muon transport. Let us consider the generic case where the initial state, $\mathbf{s}_i$, is at a vertex right after a \ac{DEL} occured and the final state, $\mathbf{s}_f$, is at the next interaction vertex, before the following \ac{DEL} occurs. In the forward \ac{MC} procedure the muon transport between these two interaction vertices is approximated by a succession of steps over which the muon energy and direction are varied due to \ac{CEL}, multiple scattering or an external magnetic field. Performing an exact inversion of this stepping process would be very inefficient, because e.g. the length of each transport step usually depends on its initial state. However the initial state is not known {\it a priori} in a backward simulation. Furthermore, the stepping part of the simulation implements a discrete approximation for a continuous transport process. Therefore, it is advantageous to define instead an approximate backward stepping process between the interaction vertices. Provided that the forward stepping is properly designed it can actually be used in backward mode with little to no modification. A possible implementation is discussed in section~\ref{sec:implementation}. Presently let us consider only the Jacobian weight factor corresponding to the backward transport between the two interaction vertices. This Jacobian weight is not straightforward to derive because it depends {\it a priori} on the discretisation procedure. The asymptotic result in the continuous limit is however simple and it is given hereafter, in lemma~\ref{lm:transport_weight}. A proof is provided in \ref{sec:proof_2}. Note that in the particular case where there is no \ac{CEL}, i.e. $\beta = 0$, lemma~\ref{lm:transport_weight} remains valid setting $\beta(\epsilon_i)/\beta(\epsilon_f) = 1$ in equation~(\ref{eq:transport_weight}).

\begin{lemma}
\label{lm:transport_weight}
Let us consider a Monte-Carlo transport process satisfying to hypotheses~\ref{hyp:base} and~\ref{hyp:direct_inverse} with total interaction length $\lambda > 0$ and continuous energy loss $\mathcal{B}$, both in units of column density. Let $\mathbf{s} = \{\epsilon, \vec{r}, \vec{u}\}$ be a Monte-Carlo state and $\rho(\vec{r})$ the variable medium density. Then, the backward Monte-Carlo weight for the transport between two interaction vertices does not depend on the details of the propagation but only on the end state properties, as:
\begin{equation}
\label{eq:transport_weight}
\omega_{i,f} = \frac{\rho(\vec{r}_f)}{\rho(\vec{r}_i)} \frac{\lambda(\epsilon_i)}{\lambda(\epsilon_f)} \frac{\beta(\epsilon_i)}{\beta(\epsilon_f)}
\end{equation}
with $\epsilon_{i,f}$ and $\vec{r}_{i,f}$ the kinetic energy and corresponding vertex position after the previous interaction or before the next interaction. $\beta(\epsilon)$ is the randomised continuous energy loss between the two vertices.
\end{lemma}

\subsection{Boundaries}
Let us now consider the case where the \ac{MC} particle hits a medium boundary or a user specified limit during its backward travel to the previous interaction vertex. Applying the backward weight given by lemma~\ref{lm:transport_weight} would not be correct in this case. The same holds when the \ac{MC} particle starts a backward propagation from a boundary or a final state, since the final position is no more an interaction vertex. A proper way to handle these boundary effects is to consider that one actually simulates a flux of particles, not a density. Thus, the initial or final state densities are defined on a surface rather than over a volume. Lemma~\ref{lm:flux_to_density}, hereafter, provides a simple and generic way to convert a surface boundary to a volume boundary or reciprocally. Note that the result is still correct in the particular case of a null \ac{CEL}. A proof is given in \ref{sec:proof_3}.
\begin{lemma}
\label{lm:flux_to_density}
Let us consider a Monte-Carlo transport process with total interaction length $\lambda > 0$ and continuous energy loss $\mathcal{B}$, both in units of column density. Let $\mathbf{s} = \{\epsilon, \vec{r}, \vec{u}\}$ be a Monte-Carlo state and $\phi_S$ the state density on an interface. Then, under hypotheses~\ref{hyp:base} and~\ref{hyp:direct_inverse}, one can arbitrarily start or stop the particle transport on the interface using an equivalent volume state density, $\rho_S$, as:
\begin{equation}
\rho_S(\epsilon, \vec{r}, \vec{u}) = \phi_S(\epsilon, \vec{r}, \vec{u}) \frac{\rho(\vec{r})}{\lambda(\epsilon)}
\end{equation}
with $\rho(\vec{r})$ the variable medium density. The corresponding transport weight is given by lemma~\ref{lm:transport_weight} substituting the initial or final state kinetic energy and position by the values at the interface crossing.
\end{lemma}

\subsection{Vertex sampling}
At an interaction vertex the \ac{MC} particle's energy and direction change discontinuously while its position remains unchanged. Let us call vertex transform the corresponding process in order to distinguish it from the transport. Inverting the vertex transform depends strongly on the nature of the process involved thus we will not give a general procedure here. Some examples are given in section~\ref{sec:implementation}. If there is no straightforward direct inversion it is preferable to rely on corollary~\ref{cor:backward_prior_sampling}.

However, as a general result, let us point out that the backward weight at the vertex transform depends only on the energies of the initial and final states. Therefore, there is no need to compute the Jacobian for the change of direction. A formal statement is given by lemma~\ref{lm:vertex_weight} hereafter and a proof in~\ref{sec:proof_4}.

\begin{lemma}
\label{lm:vertex_weight}
Let there be a discrete vertex transform from state $\mathbf{s}_i = \{\epsilon_i, \vec{r}, \vec{u}_i\}$ to state $\mathbf{s}_f = \{\epsilon_f, \vec{r}, \vec{u}_f\}$. Let the corresponding differential cross-section be invariant by rotation. Then, the backward weight of the vertex transform, $\omega_{i,f}$, depends only on the initial and final kinetic energies as:
\begin{equation}
\omega_{i,f} = \left| \frac{\partial \epsilon_i}{\partial \epsilon_f} \right|
\end{equation}
\end{lemma}

\subsection{Flux sampling}
A generic backward sampling procedure for a stationary flux of \ac{MC} particles can be derived by combining the previous results. Indeed, when both initial and final conditions are specified on a surface the backward weights given by lemmas~\ref{lm:transport_weight} and~\ref{lm:flux_to_density} nicely simplify out. Thus the final recipe is analogous to the \ac{AMC} procedure.

\begin{corollary}
\label{cor:backward_flux}
Let there be a Monte-Carlo transport process with interaction length $\lambda > 0$ and continuous energy loss $\mathcal{B}$, both in units of column density. Let us assume that the vertex transforms can be backward sampled, directly or using a bias distribution. Let us write $\omega_b$ the corresponding bias weight if any. Under hypotheses~\ref{hyp:base} and~\ref{hyp:direct_inverse}, a backward Monte-Carlo estimate of the flux of a final state $\mathbf{s}_f = \{\epsilon_f, \vec{r}_f, \vec{u}_f\}$ is given by backward propagating $N$ particles from $\mathbf{s}_f$ up to an initial flux boundary $\phi_i(\epsilon_i, \vec{r}_i, \vec{u}_i)$, with the following set of rules:
\begin{enumerate}[(i)]
	\item{The candidate interaction vertices are drawn according to equation~(\ref{eq:interaction_backward_kinetic}) or any equivalent procedure.}
	\item{Any backward transport from the kinetic energy $\epsilon_j$ to $\epsilon_{j+1} \geq \epsilon_{j}$ generates an effective backward weight:
	\begin{equation}
	\omega_{j+1,j}^{(T)} = \frac{\beta(\epsilon_{j+1})}{\beta(\epsilon_j)}
	\end{equation}
	whatever the start or end condition at $\mathbf{s}_j$ and $\mathbf{s}_{j+1}$ are, where $\beta(\epsilon)$ is the realised \ac{CEL} on the path.}
	\item{At any interaction vertex the effective backward weight for the interaction is: 	
	\begin{equation}
	\label{eq:vertex_weight}
	\omega_{j+1, j}^{(V)} = \frac{\lambda(\epsilon_j)}{\lambda(\epsilon_{j+1})} \left| \frac{\partial \epsilon_{j+1}}{\partial \epsilon_j} \right| \omega_b(\mathbf{s}_{j+1}; \mathbf{s}_j)
	\end{equation}
	with $\mathbf{s}_j$ and $\mathbf{s}_{j+1}$ the states after and before the interaction, such that $\epsilon_{j+1} \geq \epsilon_{j}$. Note that if a bias procedure is used for the backward sampling of the interaction, the Jacobian factor $|\partial \epsilon_{j+1} / \partial \epsilon_j|$ refers to the bias process. Otherwise one must set $\omega_b=1$.}
\end{enumerate}
Let $\omega_{i,k}$ be the total backward weight given by the previous rules for the $k^\text{th}$ Monte-Carlo event, and $\mathbf{s}_{i,k} = \{\epsilon_{i,k}, \vec{r}_{i,k}, \vec{u}_{i, k}\}$ be the corresponding initial state. Then a backward estimate of the stationary flux of the final state is:
\begin{equation}
\phi_f(\epsilon_f, \vec{r}_f, \vec{u}_f) \simeq \frac{1}{N} \sum_{k=1}^N{w_{i,k} \phi_i(\epsilon_{i,k}, \vec{r}_{i,k}, \vec{u}_{i, k})}
\end{equation}
\end{corollary}

Another obvious corollary concerns the \ac{BMC} integration over the final states of a stationary flux. It is stated hereafter as corollary~\ref{cor:integral_flux}. Note that the integration might concern only a subset of attributes, for example only the kinetic energy thus yielding the integral flux instead of the rate.
\begin{corollary}
\label{cor:integral_flux}
Let us consider the conditions of corollary~\ref{cor:backward_flux} with a bias distribution, $\phi_b(\mathbf{s}_f)$, for the final state over a domain $\mathcal{D}$. Then, a backward Monte-Carlo estimate of the rate of the final flux over $\mathcal{D}$ is given as:
\begin{equation}
\int_\mathcal{D}{\phi_f d\mathbf{s}_f} \simeq \frac{1}{N} \sum_{k=1}^N{w_{i,k} \frac{\phi_i(\mathbf{s}_{i,k})}{\phi_b(\mathbf{s}_{f,k})}}
\end{equation}
with $\omega_{i,k}$ computed according to corollary~\ref{cor:backward_flux} and where the Monte-Carlo final states $\mathbf{s}_{f,k}$ are generated over $\mathcal{D}$ according to $\phi_b$.
\end{corollary}

\subsection{Multiple discrete processes}
So far it was convenient to consider the \ac{DEL} as a single process with a total interaction length $\lambda$. Following this scheme, in the forward case whenever an interaction occurs one must first randomise the sub-process responsible for the interaction. This is done according to the occurrence probability, $p_\ell$, as:
\begin{equation}
p_\ell(\epsilon_i) = \frac{\lambda(\epsilon_i)}{\lambda_\ell(\epsilon_i)}
\end{equation}
where $\lambda_\ell$ is the interaction length for the $\ell^\text{th}$ process, with $1/\lambda = \sum_{\ell=1}^m{1/\lambda_\ell}$, and where $\epsilon_i$ is the kinetic energy before the interaction. In the backward case $\epsilon_i$ is not known {\it a priori}. Hence one needs to rely on a bias for the sub-process selection. For example, one can rely on corollary~\ref{cor:backward_mixture_sampling} and draw the process according to $p_\ell(\epsilon_f)$, with $\epsilon_f$ the kinetic energy after interaction. This will be a good selection procedure provided that $p(\epsilon_i) \simeq p(\epsilon_f)$ which is likely to be correct for muon transport since $\epsilon_f \simeq \epsilon_i$ for most \ac{DEL} occurrences. A natural way to perform this biasing, in close analogy with the forward \ac{MC}, is described in lemma~\ref{lm:multi_process} hereafter. A proof is given in~\ref{sec:proof_5}.

\begin{lemma}
\label{lm:multi_process}
Let there be a Monte-Carlo transport process with $m$ independent discrete energy loss sub-processes with interaction lengths $\lambda_{\ell=1,...,m} > 0$ and a continuous energy loss $\mathcal{B}$, all in units of column density. Let us write $\omega_b^{(\ell)}$ the bias weight for the randomisation of the vertex transform for process $\ell$, if any biasing is used. Under hypothesis~\ref{hyp:base} and assuming that $\sigma_\ell = 1/\lambda_\ell > 0$ for all sub-processes,  the following backward procedure is a correct simultaneous sampling of the previous hard interaction vertex and of its corresponding sub-process:
\begin{enumerate}[(i)]
{\item For all $\ell$, draw $\epsilon_{i,\ell}$ according to equation~(\ref{eq:interaction_backward_kinetic}) but substituting $\lambda$ with $\lambda_\ell$.}
{\item Select $\epsilon_{i,\ell_0} = \min(\epsilon_{i,\ell})$ as the kinetic energy of the previous vertex, after interaction, and consider the corresponding sub-process, $\ell_0$, as the interacting one.}
{\item At any interaction vertex, the vertex weight, $\omega^{(V)}_{j+1,j}$, for this backward sampling procedure is given by equation~(\ref{eq:vertex_weight}) but substituting $\lambda$ with $\lambda_{\ell_0}$ and $\omega_b$ with $\omega_b^{(\ell_0)}$.}
\end{enumerate}
\end{lemma}

Note that lemma~\ref{lm:multi_process} requires the macroscopic cross-sections of all sub-processes to be not null, $\sigma_\ell > 0$, not only the total one as stated in hypothesis~\ref{hyp:direct_inverse}. This is not rigorously true for a practical implementation, as discussed in section~\ref{sec:implementation} where another strategy is detailed. For example, the cross-section for a given sub-process could be set to zero below a threshold value of the kinetic energy, $\epsilon_0$. In this case the present bias procedure would yield a null probability to originate from the corresponding sub-process for all particles with $\epsilon_f < \epsilon_0$. This is obviously incorrect since a particle could have $\epsilon_i > \epsilon_0$ before the interaction but $\epsilon_f < \epsilon_0$ after.

\subsection{Adjoint Monte-Carlo}
Although derived using different methods, the \ac{BMC} procedure given by corollaries~\ref{cor:backward_flux},~\ref{cor:integral_flux} and lemma~\ref{lm:multi_process} and the \ac{AMC} derived by~Desorgher~{\it et al.}~\cite{Desorgher:2010} are almost identical in practice. The only difference concerns the randomisation of the vertex transform and the corresponding weight, $\omega^{(V)}_{j+1,j}$.  Using the adjoint process as an approximation of the inverse process is a particular biasing procedure for backward randomising the vertex transform. This particular biasing procedure leads directly to the results of Desorgher~{\it et al.}~\citep{Desorgher:2010} as shown formally in lemma~\ref{lm:adjoint_montecarlo}, for which a proof is provided in \ref{sec:proof_6}. Note however that for this method to be efficient one must postulate a relevant initial state density, as a bias factor $h_b(\epsilon_f; \epsilon_i)$, when computing the adjoint cross-section. Desorgher {\it et al.} set $h_b(\epsilon_f; \epsilon_i) = \epsilon_f/\epsilon_i$ which is also good for muon transport since the \ac{DCS} are soft, i.e. it is unlikely that the initial state has a much larger kinetic energy than the final one. In addition one must set an {\it a priori} high cut kinetic energy, $\epsilon_\text{max}$, above which no muon might exist. In this case the backward procedure is more flexible and directly well conditioned. If no simple direct inverse exists, one can bias with an analytically convenient approximation of the forward \ac{DCS}, easy to invert. Thus, there is no restriction on the primary muon energy. In addition, there is no need to duplicate the pre-computations and tabulations required for the adjoint cross-sections. Note also that for efficiency, the adjoint path lengths of all sub-processes are rescaled to the forward ones in Desorgher's~{\it et al.} implementation. This is not correct when the forward cross-section of a sub-process is null, which happens in practice, for example below an energy threshold. As a result the corresponding inverse processes are arbitrarily suppressed. The \ac{AMC} method however offers a conceptually straightforward inversion scheme since it is analogue to the forward case. But this comes at the price of obfuscating the flexibility of the backward scheme.

\begin{lemma}
\label{lm:adjoint_montecarlo}
Let us consider a discrete energy loss process with interaction length $\lambda$. Let us define the following integral:
\begin{equation}
F_{\sigma, h_b}^\dagger(\epsilon_i; \epsilon_f) = \int_{\epsilon_f}^{\epsilon_i}{\frac{\partial \sigma}{\partial \epsilon_f}(\epsilon_f; \epsilon_i) h_b(\epsilon_i, \epsilon_f) d\epsilon_i}
\end{equation}
where $\sigma = 1/\lambda$ is the macroscopic cross-section and with $h_b > 0$ a unit less bias function depending on $\epsilon_i$ and $\epsilon_f$, e.g. $h_b(\epsilon_i, \epsilon_f) = \epsilon_f/\epsilon_i$. If $\lambda^\dagger = 1/F_{\sigma, h_b}^\dagger(+\infty; \epsilon_f) > 0$, then:
\begin{equation}
\label{eq:adjoint_cdf}
P^\dagger_{\sigma, h_b}(\epsilon_i; \epsilon_f) = F_{\sigma, h_b}^\dagger(\epsilon_i; \epsilon_f)/F_{\sigma, h_b}^\dagger(+\infty; \epsilon_f) 
\end{equation}
is a valid bias cumulative distribution function for the inverse vertex transform, i.e. for drawing the initial kinetic energy, $\epsilon_i$, given the final one, $\epsilon_f$. The corresponding vertex weight, defined in corollary~\ref{cor:backward_flux} or lemma~\ref{lm:multi_process}, is:
\begin{equation}
\label{eq:adjoint_weight}
\omega_{i, f}^{(V)} = \frac{\lambda(\epsilon_f)}{\lambda^\dagger(\epsilon_f)} \frac{1}{h_b(\epsilon_i, \epsilon_f)}
\end{equation}
\end{lemma}

\section{Implementation}
\label{sec:implementation}
The \ac{BMC} procedure discussed in section~\ref{sec:backward_mc} has been implemented as a library initially dedicated to the transport of high energy muons used in muography. Later it was extended to taus as well. The library is named \ac{PUMAS} where the acronym reads backwards. For portability the code is written in C99 with the standard library as sole dependency. It is available from GitHub~\cite{GitHub:PUMAS} under the terms of the GNU LGPLv3 licence. \ac{PUMAS} was designed with a multi-threaded usage in mind, though it does not rely on a specific multi-threading implementation. The bulk of its used memory is read-only data relative to material properties and physics tabulations. This shared memory is dynamically preallocated by the user with a critical initialisation call. Afterwards, each simulation thread implements its own context and consumes a negligible amount of memory. The memory footprint of \ac{PUMAS} is indeed small, $300$~kB for the library code plus $100$ to $200$~kB per loaded material. Additionally, calls to the standard library require a few extra MB. Thus, \ac{PUMAS} parallelisation is very likely to be only CPU-bound on modern hardware. With a proper parallel framework, e.g. using~HPCSim~\cite{GitHub:HPCsim}, a linear scaling of the processing speed with the number of available CPU cores can be achieved as was shown by Schweitzer~\cite{Schweitzer15}.

In the following, we start with a brief overview of the forward \ac{MC} procedure that has been implemented in \ac{PUMAS}. Then, we discuss the modifications for the backward case. Note that for the sake of simplicity we limit the discussion to muons and muography, although \ac{PUMAS} can also handle taus.

\subsection{Energy loss}
A key property for muography is the muon average energy loss, $\left< dE/dx \right>$, including both \ac{CEL} and \ac{DEL}. Within the \ac{CSDA}, the average energy loss governs the observed rate of transmitted muons through rocks in most practical cases, i.e. below $\simeq 1$~km rock depth. Tabulations of muon's $\left< dE/dx \right>$ are provided online by the \ac{PDG}~\cite{PDG:2014, PDG:AtomicProperties15} for a very broad range of materials. Therefore, \ac{PUMAS} was designed to take these tabulations as numerical input. The \ac{PDG} tables provide the $\left< dE/dx \right>$ split between the four energy loss processes relevant for muons, i.e. ionisation, pair production, bremsstrahlung and photonuclear interactions. Using these tabulated data, \ac{PUMAS} allows three levels of detail for the simulation of the energy loss, configurable per simulation context as follows.
\begin{enumerate}[(i)]
\item{\ac{CSDA} : the energy loss is fully deterministic with only \ac{CEL}, i.e. \mbox{$\beta = \left< dE/dx \right>$} and $\lambda = +\infty$.}
\item{{\it hybrid} : the energy loss is split between a deterministic \ac{CEL}, \mbox{$\beta < \left< dE/dx \right>$}, and stochastic catastrophic \ac{DEL}, $0 < \lambda < +\infty$.}
\item{{\it detailed} : the energy loss is split as previously but the ionisation component of the \ac{CEL} also has small stochastic fluctuations.}
\end{enumerate}
For the {\it hybrid} and {\it detailed} cases the splitting between \ac{CEL} and \ac{DEL} is fixed at a relative cut value on the transferred energy of $\nu_{CEL}=5\%$. This might seem high but it was shown by Sokalski {\it et al.}~\cite{MuM01} that values as high as $10$~\% can reasonably be used for high energy muons. The value set in \ac{PUMAS} was cross-checked to be a good compromise between speed and accuracy for muography applications. As a result the {\it hybrid} simulation scheme {\it \`a la} MUM is very efficient.

For the simulation of \ac{DEL}, the {\it hybrid} and {\it detailed} cases require explicit \ac{DCS} for each  process. The bremsstrahlung \ac{DCS} has been implemented following Groom {\it et al.}~\cite{Groom01}, i.e. the same as was used for computing the \ac{PDG} tables. However, for pair production Geant4's implementation~\cite{Geant4,Geant4:PRM} is used instead, since it is numerically much simpler and provides identical results. Also, for photonuclear interactions the DRSS \ac{DCS}~\cite{DRSS} is used rather than the Bezrukov and Bugaev \ac{DCS}~\cite{Bezrukov_Bugaev:1981}, used also by Groom {\it et al.}. Indeed, the latter cross-section underestimates the energy loss at PeV energies and above. Though the impact is mild for muography, it matters when considering ultra relativistic taus. Finally, for ionisation it is enough to consider only close interactions for \ac{DEL}. The corresponding knock-on electron cross-section, including a high energy radiative correction, is modelled according to MUM~\cite{MuM01}.

At \ac{PUMAS} initialisation the average energy loss of all \ac{DEL}, $\left< dE/dx \right>_{DEL}$, is computed by numerical integration of the \ac{DCS}. Then, the average value of the \ac{CEL} is rescaled as $\left< \beta \right> = \left< dE/dx \right>_{PDG}-\left< dE/dx \right>_{DEL}$, where $\left< dE/dx \right>_{PDG}$ is the average energy loss read from the \ac{PDG} tables. Note however that these tabulations have been initially computed only up to $10^5$~GeV and are not accurate above even though higher values are reported. Indeed, radiative losses are linearly extrapolated above $10^5$~GeV and hence underestimated. In rocks the underestimation is of the order of $\simeq 5\%$ at $10^5$~GeV. Of course alternative $\left< dE/dx \right>$ tables might be used as input as long as they have a format similar to the \ac{PDG} ones. In particular, \ac{PUMAS} includes an executable that generates energy loss tables, in the \ac{PDG} format, according to its own \ac{DCS}. In addition, since the initialisation of \ac{PUMAS} can take a few seconds the library can also dump and reload a binary summary of all its static data, hence allowing to skip the initial pre-computations.

The {\it detailed} case has been implemented following PENELOPE~\cite{Baro:1995, Sempau:1997}. The \ac{CEL} is fluctuated around its mean value, $\left< \beta \right>$, using a truncated Gaussian distribution for large steps, a uniform distribution for intermediate steps and a mixture \ac{pdf} for short step lengths, as given in section~4.2 of PENELOPE's 2001 manual~\cite{PENELOPE:2001}. In the continuous limit, the straggling parameter $\Omega^2(\epsilon)$ is given by the energy loss variance in a thick absorber, see e.g~\cite{ICRU49, Geant4:PRM}. In order to improve the accuracy for large steps and for consistency with the backward case, $\frac{1}{2}(\Omega^2(\epsilon_i)+\Omega^2(\epsilon_f^*))$ is used instead of $\Omega^2(\epsilon_i)$. Thus, one performs a trapeze discrete integral of the straggling along the step. In previous expressions, $\epsilon_i$ stands for the kinetic energy at step's start and $\epsilon_f^*$ is the kinetic energy at step's end in the ${\it hybrid}$ case, i.e. for mean \ac{CEL} $\left< \beta \right>$. Note that the effect of this soft straggling is small : for GeV muons it leads to a $\sim 5$~\% fluctuation on the muon range.

For the transport, the four inelastic \ac{DEL} are treated as a single process with total interaction length $\lambda_\text{in}$. In the {\it hybrid} case the kinetic energy at the next interaction is randomised {\it a priori} from equation~(\ref{eq:forward_MUM}). Then, the corresponding column density can be derived in a unique way. In the {\it detailed} case this is not efficient. Instead the rejection method described in section~4.3.1 of PENELOPE's manual is used at each transport step. At an interaction vertex, a sub-process is selected and the muon final kinetic energy, $\epsilon_f$, must be randomised according to the corresponding differential cross-section. Above $10$~GeV, considering the high cut value $\nu_{CEL}$ used, the \ac{DCS} are monotone decreasing functions of the energy transfer. Thus, a Zigourat algorithm~\citep{Marsaglia2000} is used to efficiently sample $\epsilon_f$ by rejection sampling from pre-computed envelopes. In addition, in order to speed up their evaluation, the \ac{DCS} are initially fitted by polynomials using a log-log mapping. The order of the polynomials was set so that an agreement better than $0.3$~\% is achieved over the range of use. Only in the case of ionisation the exact \ac{DCS} for the knock-on electrons is computed on the flight, since its expression is already very light CPU-wise. Below $10$~GeV the sampling of $\epsilon_f$ is biased using a power law and the Monte-Carlo weights are corrected accordingly. Consequently, even in the forward case, \ac{PUMAS} is not rigorously an analog Monte-Carlo.

\subsection{Transverse transport}
\ac{TT} can be enabled or disabled per simulation context. This allows useful optimisations since the impact of \ac{TT} on muography is negligible for high energy muons, above $\simeq 100$~GeV, while the simulation of \ac{TT} is very CPU consuming. For \ac{TT}, in addition to the four energy loss processes one has to consider the Coulomb scattering off screened nuclei. It is the dominant source of \ac{TT} though its contribution to the energy loss is negligible. An accurate estimation of the Coulomb differential cross-section is given by Salvat~\cite{Salvat13}, for protons, using the Eikonal approximation. The application to muons is straightforward. However, the Eikonal computation is too CPU intensive to be directly used in \ac{MC}. Therefore, Salvat relies on detailed tabulations. Instead, in \ac{PUMAS} the Coulomb scattering has been implemented following Salvat~\cite{Salvat13} but substituting an approximate result to the Eikonal estimate, i.e. a Wentzel distribution. The atomic screening parameter is taken from Kuraev et al.~\cite{Kuraev:14}. It includes high energy Coulomb corrections to Moliere's first estimate. In addition, the finite nucleus size, relevant for high energy muons, is taken into account by a nuclear form factor. Following ELSEPA~\cite{Salvat05}, the nucleus is modelled by Helm's uniform-uniform distribution. This allows analytical integration while being accurate enough for the present purpose. The final result is a fair trade off between speed, memory usage and accuracy. It yields the correct tail behaviour while being accurate to a few percent on the standard deviation, for relativistic muons. 

As for the energy loss, the \ac{TT} is split between a continuous multiple scattering component and catastrophic discrete events. For Coulomb interactions the split is done by cutting on the deflection angle following Fern\'andez-Varea {\it et al.}~\cite{Salvat93}. The corresponding interaction length, $\lambda_\text{el}$, is tabulated at initialisation. Hence, a discrete Coulomb event is handled as a second process distinct from the \ac{DEL}. The deflection angle at the vertex of a Coulomb event is randomised with the inverse \ac{cdf} method. The implemented process has an analytical expression for the \ac{cdf}. The inversion is done by a bracketing algorithm, initialised with an approximate solution. In the {\it hybrid} and {\it detailed} energy loss cases the deflection angle at the vertex of a \ac{DEL} event is randomised following Geant4's implementation. Finally, the multiple scattering process is simulated using a truncated Gaussian distribution of width $\sigma_\theta(\epsilon)$. The dominant contribution arises from Coulomb scattering. The soft part of photonuclear interactions is also included in $\sigma_\theta(\epsilon)$. Other soft processes are neglected. For accuracy and consistency with the backward case, the multiple scattering width is integrated over \ac{MC} transport steps, as in the case of the energy straggling. Hence, $\frac{1}{2}(\sigma^2_\theta(\epsilon_i)+\sigma^2_\theta(\epsilon_f))$ is used instead of $\sigma^2_\theta(\epsilon_i)$ or $\sigma^2_\theta(\epsilon_f)$.

\subsection{Backward tweaks}
For deterministic \ac{CEL}, a convenient backward transport process is to simply use the forward process with the substitution $\vec{u} \rightarrow -\vec{u}$, i.e. the propagation direction is inverted. Since the forward transport procedure was designed accordingly, the accuracy of the corresponding backward transport is similar to the forward one. This is why it was important to integrate the multiple scattering over the transport steps. Using only the initial or final kinetic energy of the step to randomise the multiple scattering would lead to an asymmetric result and less accurate forward or backward multiple scattering. In the {\it detailed} case, for small fluctuations a good approximation for the straggling of the energy loss is to use the forward algorithm by exchanging $\epsilon_i$ and $\epsilon_f$. This is obviously wrong since one uses $\epsilon_i^*$ instead of $\epsilon_i$ which is not known in a backward step. But in case of large steps where the relative straggling is small, or if several transport steps are summed, the effect is negligible. The main problem is actually to correctly estimate the backward weight. When considering small fluctuations of the energy loss, a good approximation for the backward transport weight between two vertices, boundaries, {\it etc}~... is given as:
\begin{equation}
\omega_{j+1,j}^{(T)} = \frac{\left<\beta\right>(\epsilon_{j+1})}{\left<\beta\right>(\epsilon_j)}
\end{equation}
Provided that the stepping is long enough, the fluctuations of the energy loss at end points are negligible with respect to the bulk variation. As shown in section~\ref{sec:validation} this is accurate enough for most cases considered in muography, resulting in discrepancies on the muon flux smaller than $1$~\%. Thus, in practice the same backward weight is used in the {\it hybrid} and {\it detailed} cases. But in the first case $\epsilon_{j+1}$ is deterministic, given $\epsilon_j$, while in the second it is stochastic.

The backward sampling of \ac{DEL} or of discrete Coulomb events is straightforward. Note that in the {\it detailed} case the rejection method used in PENELOPE and PUMAS is identical to its inverse. A difficulty arises however from the fact that for numerical convenience the total inelastic cross-section, $1/\lambda_\text{in}$, is set to zero below a threshold value $\epsilon_{0} > 0$, corresponding to the smallest non null tabulated value. In addition, even for $\epsilon > \epsilon_0$, for some sub-processes the cross-section is also set to zero for small kinetic energy values. For example, the parametrisations for the computation of the photonuclear cross-section become unstable at low energies. It follows that the backward sampling method discussed in section~\ref{sec:backward_mc} would fail for inelastic processes. Consequently, some possible cases occuring during the forward sampling would be missed in the backward sampling.

Instead, the following procedure has been implemented, relying on corollaries~\ref{cor:backward_prior_sampling} and~\ref{cor:backward_mixture_sampling}. The total inelastic cross-section is regularised by adding a virtual {\em do nothing} sub-process for $\epsilon \leq \epsilon_0$ with a large interaction length, set equal to the last non null tabulated value. The backward sampling of \ac{DEL} at vertices proceeds in three steps, as following.
\begin{enumerate}[(i)]
{\item First, a backward {\it do nothing} process might occur with a probability $p_0$, set as:
\begin{equation}
p_0 = \frac{\ln(\frac{1}{\nu_\text{CEL}})}{\ln(\frac{1}{\nu_\text{CEL}})+\ln \left(\frac{\epsilon_0}{\epsilon_0-\epsilon_f} \right)} \theta (\epsilon_0-\epsilon_f)
\end{equation}
If a {\it do nothing} process is drawn nothing occurs at the vertex, except that the \ac{MC} weight is updated by multiplying it by $1/p_0$. Otherwise, an inelastic event occurs and an additional weight factor $1/(1-p_0)$ has to be applied.}
{\item Then, a bias power law distribution is assumed for the discrete energy transfer at the vertex, independently of the yet unknown inelastic sub-process, as:
\begin{equation}
\rho_\mathrm{DEL}(\epsilon_f; \epsilon_i) \propto \frac{1}{(\epsilon_i-\epsilon_f)^{\alpha_\mathcal{E}}} 
\end{equation}
Using corollary~\ref{cor:backward_prior_sampling} the initial kinetic energy, $\epsilon_i$, is backward sampled given $\epsilon_i \geq \min(\epsilon_0, \epsilon_f/(1-\nu_\text{CEL}))$.}
{\item Finally, the inelastic sub-process is drawn according to the differential cross-sections from $\epsilon_i$ to $\epsilon_f$, instead of using the integrated ones at $\epsilon_i$.}
\end{enumerate}

If \ac{TT} is enabled, given both $\epsilon_i$, $\epsilon_f$ and the sub-process, the backward randomisation of the deflection angle is identical to the forward case. Note that (i) is a direct application of corollary~\ref{cor:backward_mixture_sampling}. The probability law for $p_0$ is arbitrary. It was set by assuming a $1/\epsilon_f$ distribution for events with $\epsilon_f < \epsilon_0$ originating from inelastic processes. Indeed, particles starting with a kinetic energy well below $\epsilon_0$ are unlikely to originate directly from an inelastic \ac{DEL} since these processes are soft.

Drawing the initial kinetic energy, $\epsilon_i$, before the sub-process ensures that no possibility is left aside. For relativistic muons, a value of $\alpha_\mathcal{E} = 1.4$ was found to give good results, i.e. low sample variance for the \ac{MC}. Randomising over the differential cross-sections in (iii) is not mandatory. However it ensures that the selected sub-process has both a non null interaction length for $\epsilon_i$ and that it indeed allows a transition from $\epsilon_i$ to $\epsilon_f$. As a result one does not backward generate useless non-physical events. The final \ac{MC} weight, corrected for the biasing procedure, is guaranteed to be not null.

\section{Validation}
Let us now consider here some benchmark cases in order to validate the \ac{BMC} implementation described in section~\ref{sec:implementation}. Two simulation schemes are considered. The first one, named hybrid hereafter, is a simplified 1-dimensional propagation with a deterministic CEL. It can be compared to MUM\cite{MuM01}. The second scheme, named detailed, implements a stochastic CEL and transverse transport as discussed previously. It can be compared to Geant4~\cite{Geant4}, for what concerns muon transport. Comparisons with MUM and Geant4 are provided in order to show that the present implementation is indeed realistic. But the main goal is actually to cross-check the \ac{BMC} and forward implementations used in PUMAS.

The comparisons have been done using the tagged version~0.8 of \ac{PUMAS}. The propagation medium was set to standard rock~\cite{StandardRock} with a uniform density of $2.65\ \mathrm{g/cm^3}$, since it is a common material for most high energy muon transport codes. In addition the energy loss table has been self generated using \ac{PUMAS}.

In order to clarify the discussion let us define formaly a few criteria used for the forecoming comparisons. Let us write $\mu_n(\mathbb{I})$ the n$^\mathrm{th}$ sample moment for a subset $\mathbb{I}$ of \ac{MC} events belonging to a given population, $\mathcal{P}$.
\begin{equation}
\mu_n(\mathbb{I}) = \frac{1}{N} \sum_{i \in \mathbb{I}}{\omega_i^n}
\end{equation}
where $N$ is the total sample size and $\omega_i$ the \ac{MC} weight of the $i^\mathrm{th}$ event. A \ac{MC} estimate of the probability to belong to $\mathcal{P}$ is given by the $1^\mathrm{st}$ sample moment, i.e. $\mathrm{Prob}[\mathcal{P}] \sim \mu_1$. Let us call {\it systematic error} the difference on $\mathrm{Prob}[\mathcal{P}]$ obtained with two different simulations, in the limit of an infinite sample size. An estimate of the statistical uncertainty on $\mathrm{Prob}[\mathcal{P}]$ is \mbox{$\sigma_\mathcal{P} = \sqrt{(\mu_2 - \mu_1^2) / (N - 1)}$}. Let us call $\sigma_\mathcal{P}$ the {\it \ac{MC} error} and the ratio $\sigma_\mathcal{P} / \mu_1$ the {\it relative \ac{MC} error}.

\label{sec:validation}

\subsection{Low energy case}
To start with, let us consider the following toy geometry in order to cross-check the \ac{BMC} implementation for muons in ionisation regime, i.e. well below the muon critical energy in standard rock, $E_c = 693$~GeV.

The propagation medium is a box of standard rock of $10$~m thickness along the $z$-axis and infinite extension along the $x$ and $y$ directions. The initial muon flux, on the $z=0$~m face, has a kinetic energy uniformly distributed over $\epsilon_i \in [5,6]$~GeV and direction $u_{i,z} = \vec{u}_i \cdot \vec{e}_z \in [0.9, 1]$ uniformly distributed in solid angle as well.

Then, one is interested in computing the outgoing flux on the $z=10$~m face. In particular, let us consider the distribution of the final state kinetic energy, $\epsilon_f$, of the muon direction, $u_{f, z}$, and of its transverse deflection, \mbox{$|x_f-x_i|$}. In addition let us select only outgoing muons with kinetic energy \mbox{$\epsilon_f \in [0.1, 1]$}~GeV and direction $u_{f,z} \in [0.9,1]$. Let us write $p_s$ the corresponding probability that the outgoing muons fulfill the selection criteria.

The selection probabilities obtained with PUMAS for different settings are given in table~\ref{tab:p_det}. In the hybrid case the agreement between the forward and backward simulations is better than the statistical uncertainty of~$0.04$~\%. In the detailed case, due to the use of an approximate backward \ac{CEL} process, there is a slight discrepancy of $1$~\%, arising from low energy transport steps. Switching on and off the transverse transport does not change the discrepancy. Increasing the initial kinetic energy however decreases the forward/backward discrepancy since the impact of small \ac{CEL} decreases. The selection probability obtained with PUMAS is close to the one obtained using Geant4, though a statistically significant ($16~\sigma$) deficit of $0.8$~\% is visible. MUM result however should only be considered qualitatively. MUM was not designed to be used at such low energies. Below $10$~GeV the mean \ac{CEL} computed by MUM starts to deviate significantly from the \ac{PDG} or Geant4 tabulations.

\begin{figure}[th]
	\center
	\includegraphics[scale=0.5]{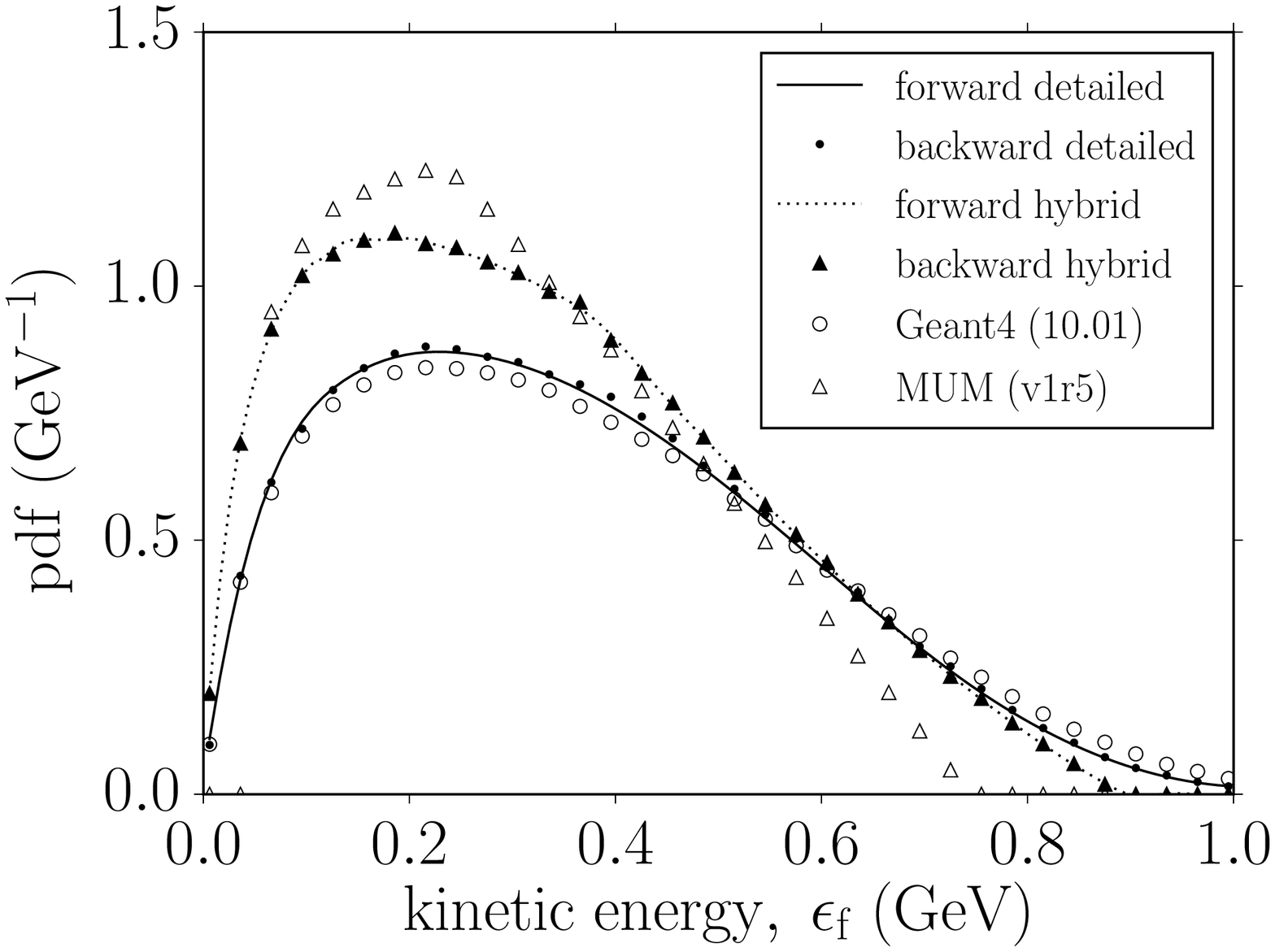}
	\caption{Distribution of the final state kinetic energy.}  \label{fig:bmc_kinetic}
	\includegraphics[scale=0.5]{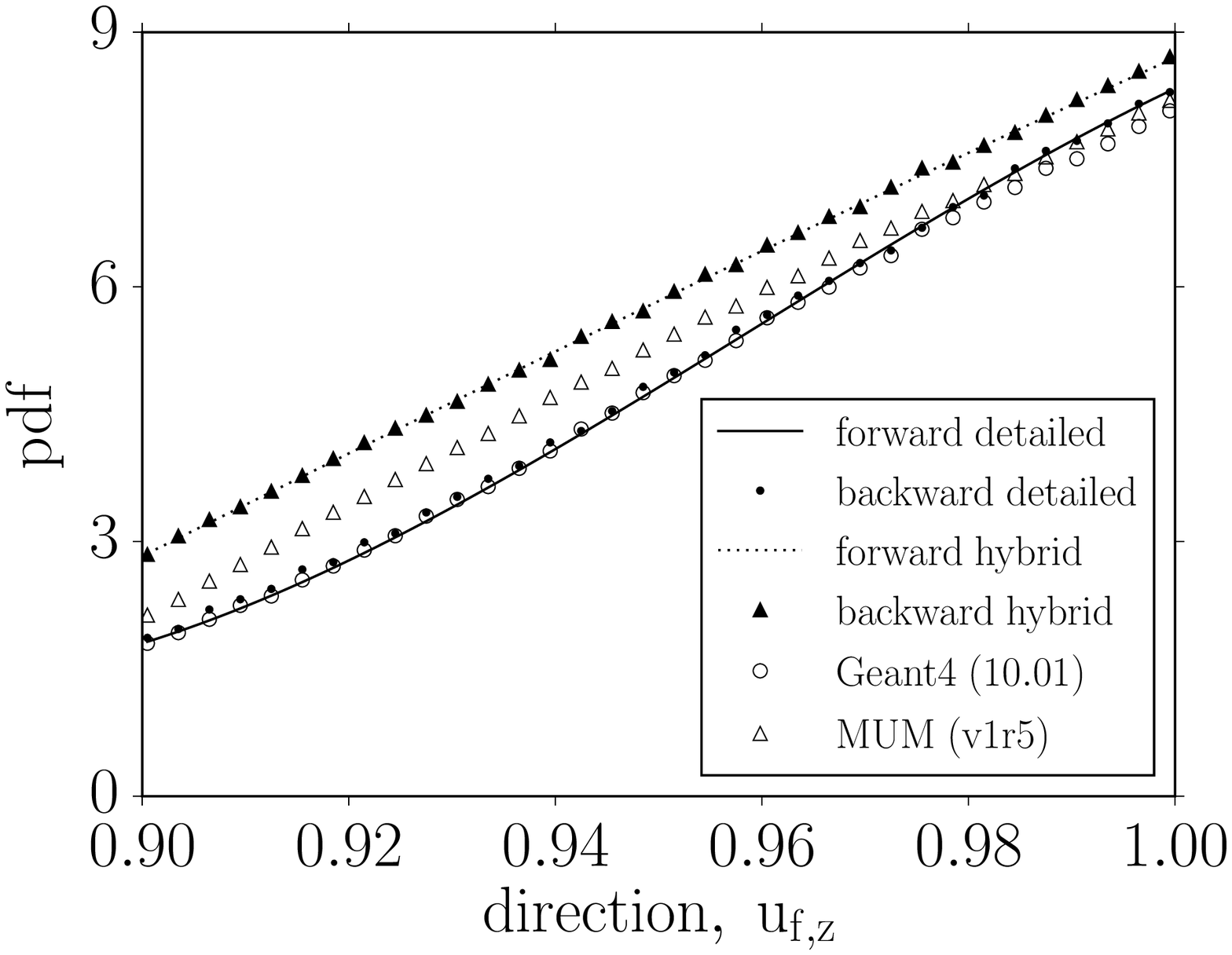}
	\caption{Distribution of the final state direction.}
\end{figure}

\begin{figure}[th] 
	\center
	\includegraphics[scale=0.5]{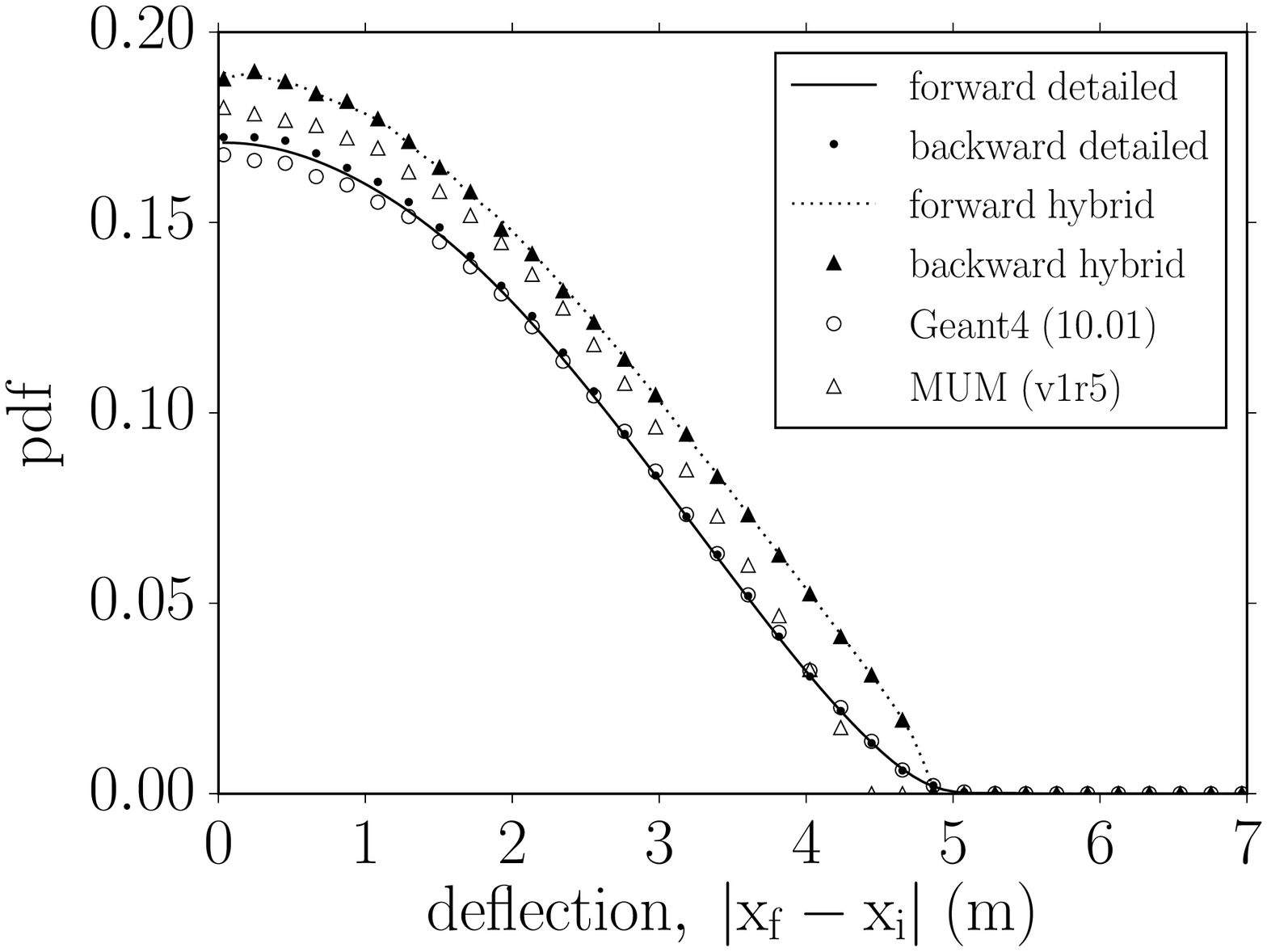}
	\caption{Distribution of the lateral deflection for selected particles.} \label{fig:bmc_deflection}
\end{figure}

Figures~\ref{fig:bmc_kinetic} to~\ref{fig:bmc_deflection} show the distribution of the selected muons. For PUMAS there is an excellent agreement on the shapes obtained with forward and backward sampling. The \ac{pdf}s agree within $1$~\%. When comparing Geant4 results to PUMAS ones the strongest discrepancy concerns the final state kinetic energy, $\epsilon_f$. Geant4 yields a $3$~\% larger average value of $\epsilon_f$ while having a slightly lower value for $p_s$. The missing events are mainly located in the forward region, for small deflections. These differences are likely due to the use of a higher cut value in PUMAS, $\nu_\text{CEL} = 5$~\%, for the splitting of the energy loss into discrete and continuous components. As a result, in this particular case PUMAS underestimates a few low energy catastrophic events that are wrongly redistributed into the \ac{CEL}. The global impact on $p_s$ is however mild and decreases as the muon initial energy increases. Hence, the high value selected for $\nu_\text{CEL}$ is of little impact for muography applications while providing an impressive CPU gain as shown in section~\ref{sec:validation-high}.

\begin{table}
\caption{Selection probability for the toy case.} \label{tab:p_det}
\begin{center}
\begin{tabular}{|c|cc|cc|} \hline \hline
   code   & sampling & scheme   & $p_s$~(\%)   & $\sigma_{p_s}$~(\%) \\ \hline \hline
   MUM    & forward  & hybrid   & 52.750 & 0.016 \\
   PUMAS  & forward  & hybrid   & 58.097 & 0.017 \\
   PUMAS  & backward & hybrid   & 58.142 & 0.037 \\ \hline
   Geant4 & forward  & detailed & 48.632 & 0.016 \\
   PUMAS  & forward  & detailed & 49.008 & 0.017 \\
   PUMAS  & backward & detailed & 49.492 & 0.034 \\ \hline \hline
\end{tabular}
\end{center}
\end{table}

\subsection{High energy case}
\label{sec:validation-high}
Let us now cross-check the high energy behaviour with a toy case taken from muography, i.e. the rate of atmospheric muons surviving after propagation through a given column density of standard rock. Let the initial spectrum of atmospheric muons, $\phi_i$, be given by Gaisser's parametrisation~\cite{Gaisser:Book, PDG:2014} for an elevation of $10$~deg and an initial kinetic energy $\epsilon_i \geq 0.1$~GeV. Note that this model is not very accurate since the parametrisation used for $\phi_i$ neglects the Earth curvature and the muon decay, which is not correct below $10$~deg of elevation nor below $0.1$~GeV. However it is precise enough for the present purpose and has the advantage of being a widely used model. Let us write $\Phi(s)$ the rate of remnant muons with kinetic energy $\epsilon_f \geq 0.1$~GeV after a path length $s$ in the rock. Note that in the detailed case the path length is integrated along the muon trajectory. Hence, at low energies it differs from the rock thickness by a detour factor (0.98 in the worst case~\cite{Groom01}). Considering the path length however allows for a direct comparison of the hybrid and detailed cases.

The muon rate as function of the path length was computed by \ac{MC} varying the path length from $1$~mm to $10$~km with a logarithmic stepping. A log-uniform bias distribution was used for the initial or final state kinetic energy. For MUM and PUMAS the initial kinetic energy was limited to $\epsilon_i \leq 10^8$~GeV. However, Geant4 does not provide a model for photonuclear interactions at such high energies. Hence, we had to limit $\epsilon_i \leq 10^6$~GeV in this case. For the hybrid case $10^8$~events were simulated per value of path length, resulting in a negligible \ac{MC} error. For the detailed case this number of events was limited to $10^7$.

\begin{figure}[th] 
	\center
	\includegraphics[scale=0.5]{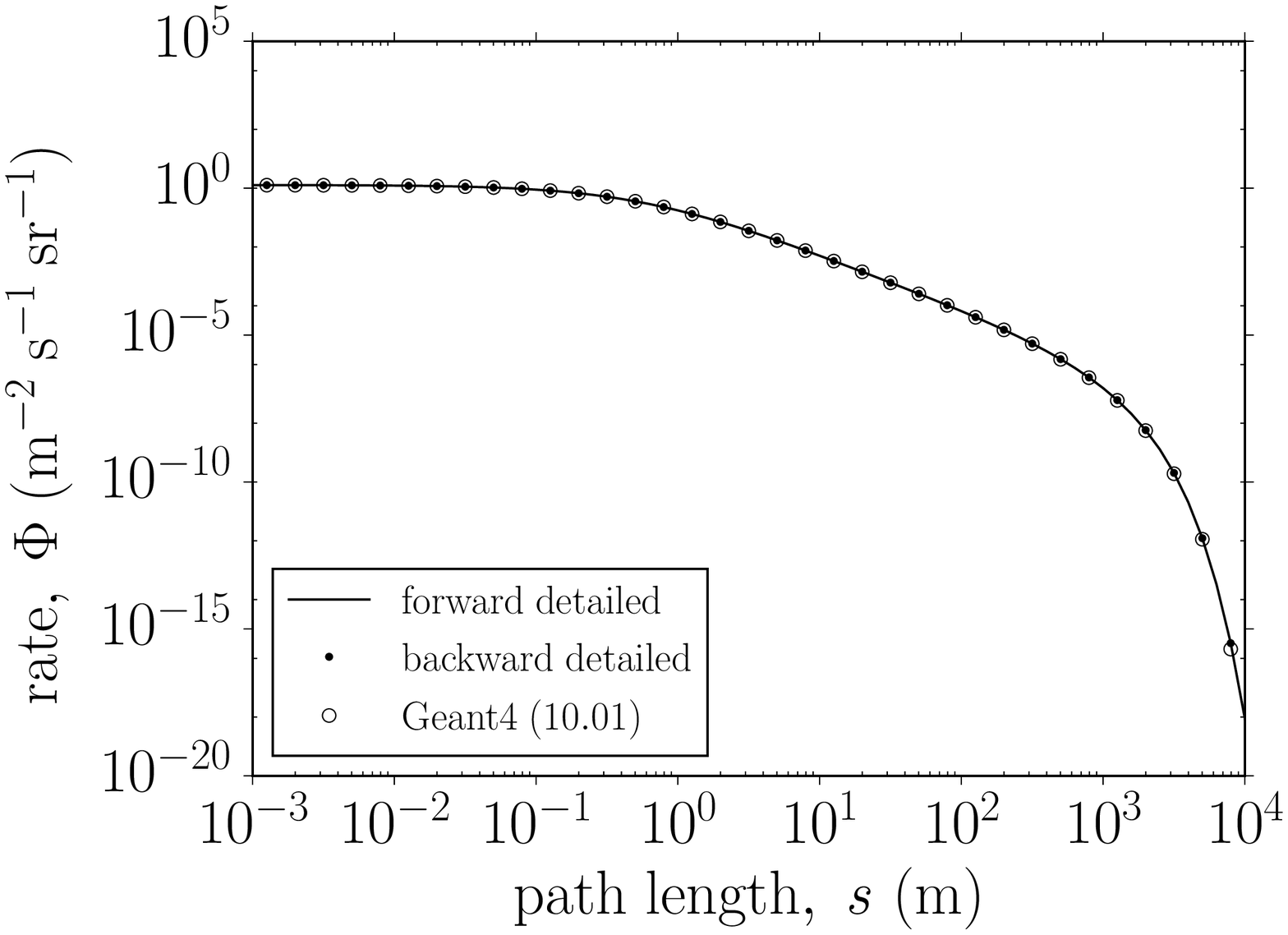}
	\caption{Integrated rate of muons transmitted through standard rock for a Gaisser spectrum at $10$~deg of elevation and extending from $100$~MeV to $1$~PeV kinetic energy.} \label{fig:prob_flux}
\end{figure}

\begin{figure}[th]
	\center
	\includegraphics[scale=0.5]{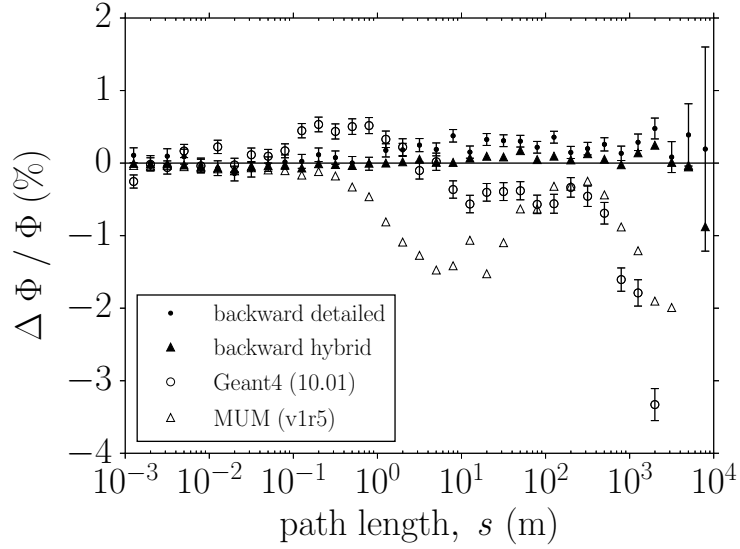}
	\caption{Relative differences on the integrated muon rate, w.r.t. the `forward hybrid` case.}  \label{fig:prob_error}
\end{figure}

The estimated rate is shown on figure~\ref{fig:prob_flux}. It spans $20$~orders of magnitude. Above $1$~km of standard rock the rate falls drastically as the range of high energy muons becomes limited by radiative processes. Hence, the \ac{MC} estimate of the rate becomes very inefficient even with a log biasing. A more detailed comparison is shown on figure~\ref{fig:prob_error} where the PUMAS forward hybrid case is taken as a reference for the comparison.

Considering only PUMAS the forward and backward results are in perfect agreement for the hybrid case, up to statistical uncertainties of $0.1$~\%. For the detailed case a $0.3$~\% difference is observed with respect to the hybrid case. The \ac{MC} errors are mostly negligible when compared to the observed difference, which can be considered as a systematic error. This is consistent with discretisation errors. Indeed, the hybrid scheme relies on a tabulation of the total number of interactions lengths, i.e. the integral in equation~(\ref{eq:survival_probability}), while the detailed case uses a tabulation of the total cross-section instead. Both are consistent only up to discrete integration errors. 

Below a few hundred meters of path length, Geant4 and PUMAS agree within $\simeq 0.5$~\% systematic effects. MUM however exhibits larger deviations, $\simeq 2$~\%, over the whole energy range investigated. This could be due to a simplified \ac{CEL}, or to differences in the interpolation of the cross-sections for \ac{DEL}s. Above a few kilometers of path length Geant4 starts to deviate significantly from PUMAS and MUM. This is consistent with uncertainties on the energy loss due to photonuclear interactions. Actually, the photonuclear cross-section is the main source of systematics on the muon range at those energies, see e.g.~Sokalski {\it et al.}~\cite{Sokalski:2001}. In addition, in this comparison Geant4 suffers from a bias due to the initial kinetic energy being limited to $10^6$~GeV, resulting in an underestimated rate for very large path lengths.

\begin{figure}[th]
	\center
	\includegraphics[scale=0.5]{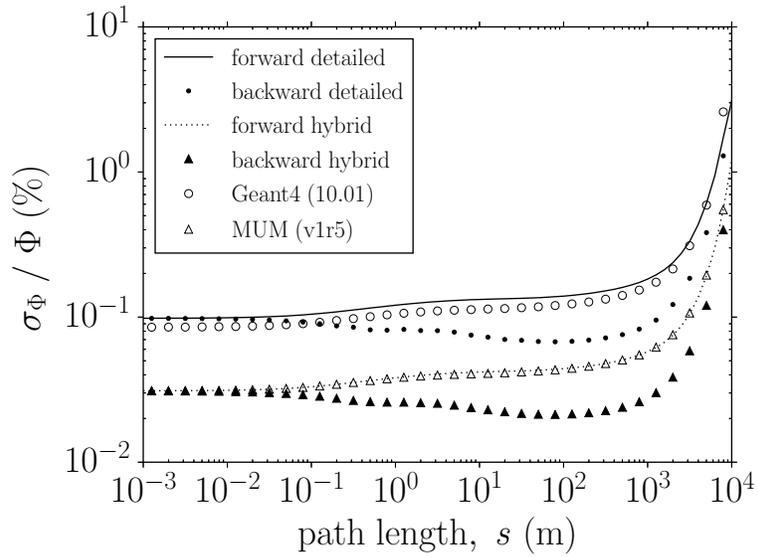}
	\caption{Relative \ac{MC} error on the integrated muon rate, $\Phi$.}  \label{fig:prob_sigma}
\end{figure}

Finally let us compare the \ac{MC} integration performances. Figure~\ref{fig:prob_sigma} shows the \ac{MC} relative error for the different simulation codes. For PUMAS, the detailed and hybrid cases lead to identical errors up to a factor $\sqrt{10}$ due to the different number of events. This was expected since the detailed and hybrid cases are compared for a same path length. Thus, the details of the transverse transport play no role. In the forward case, PUMAS and MUM accuracies are identical. However, Geant4 is slightly more efficient than PUMAS below $1$~km path lengths. This is because the muons propagated with Geant4  have a lower maximal value for their initial kinetic energy.

Considering the backward case, the \ac{MC} error is systematically better than the forward one. The improvement is marginal though in this case, up to a factor $2$ between $10$~m and $1$~km path lengths. But since the present purpose is to validate the \ac{BMC} implementation it is convenient to consider cases where the forward and backward cases are evenly efficient.

To complete this performance comparison, let us take into account the CPU cost. Let us introduce $\tau_{100}$ as the average CPU time required to reach a $1$~\% \ac{MC} error on the muon rate, pruned from any initialisation time. Of course the results depend on the hardware on which the code runs, so this is only indicative. For MUM and PUMAS the same PC was used, an Intel Xeon E5-$4620$ at $2.2$~GHz having $64$~cores with $128$~GB total memory. For Geant4, since the required CPU time was significantly higher the code was run on the CPU cluster of CC-IN2P3. It was however cross-checked that the average CPU performances of both systems agree within 10\% for Geant4.

\begin{figure}[th]
	\center
	\includegraphics[scale=0.5]{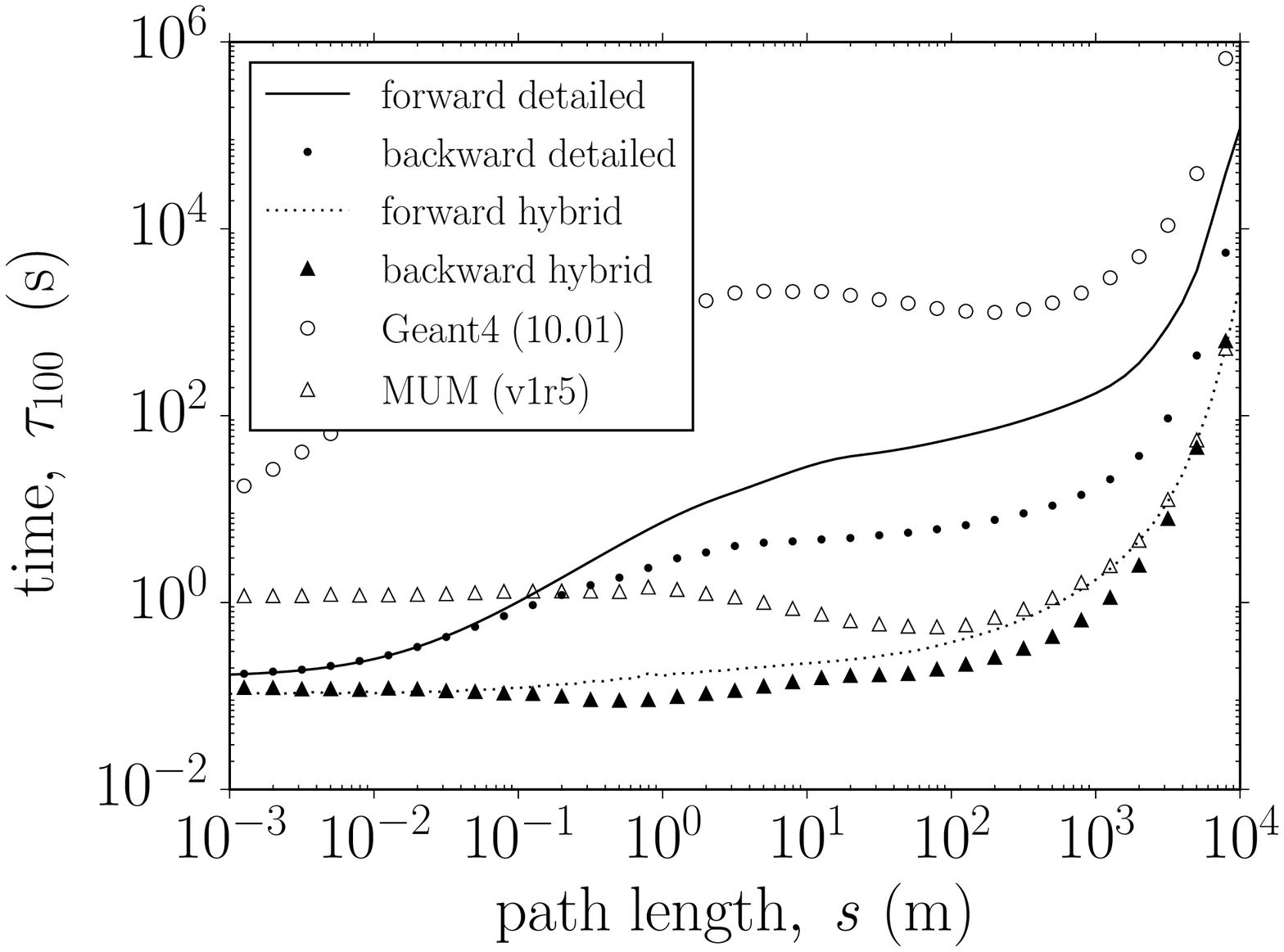}
	\caption{Processing time for a $1$~\% \ac{MC} error on the integrated muon rate, $\Phi$.} \label{fig:prob_time}
\end{figure}

A summary of the achieved performance is shown on figure~\ref{fig:prob_time}. Below $3$~km path lengths the best performance is achieved with PUMAS in backward hybrid mode. At larger depths the forward hybrid mode becomes more efficient however, e.g. by $50$~\% at $10$~km. This could be due to a non optimal backward sampling of \ac{DEL} or to an interplay between the \ac{MC} error and the CPU cost of events. Nevertheless, path lengths larger than $3$~km are not of practical interest for muography because the transmitted flux becomes ridiculously small. Note also that PUMAS forward hybrid and MUM have similar performances, within $10$~\%, when crossing more than hundred meters of rock.

Achieving such performances while being accurate enough is not straightforward. It requires a balance between float and double usage as well as efficient approximations for the differential cross-sections.

When considering the detailed simulations, PUMAS, even in the forward mode, is more efficient by at least 1 order of magnitude with respect to Geant4. In addition, for PUMAS, a significant gain is observed by using the backward mode, a factor of $10$ at large depths with respect to the forward one. In the detailed case the bottleneck is the simulation of the \ac{TT}. In the forward mode a lot of CPU is wasted in the detailed \ac{TT} of low energy events that would actually die without ever reaching large depths. Using the backward mode and switching between detailed mode at low energies and hybrid mode at high energies, e.g. above $100$~GeV, the CPU time can be decreased by $2$ to $3$ orders of magnitude with respect to Geant4.

\section{Conclusion}
\ac{BMC} provides a simple way to invert the simulation flow in a muon transport \ac{MC}. The elementary step processes are reverted one by one, when a trivial inverse exists, or using a bias strategy when it does not. This is a correct procedure provided that a Jacobian weight factor is applied at each step, in order to preserve unitarity. Backward sampling can provide drastic CPU gains by simulating only events that can be observed. In practice, it is similar to an adjoint \ac{MC} but more flexible. A proof of principle has been provided using a dedicated library implemented for muography imaging: PUMAS. The agreement between the forward and backward sampling result is better than the $0.1$~\% \ac{MC} error at high energies. At low energy, due to an approximate backward procedure for the fluctuations in the ionisation loss, systematics of $1$~\% can be observed. However, this systematic is washed out when considering that muography imaging uses atmospheric muons with a broad energy spectrum.

\appendix
\section{Proof of lemma~\ref{lm:backward_sampling}}
\label{sec:proof_1}

Let us write $Y = g^{-1}(\mathbf{s}_f; X)$ the random variable derived from $X$ and $g^{-1}$. The transformed variable has a density $\rho_Y$ given as:
\begin{equation}
\begin{split}
\label{eq:backward_density}
\rho_Y = & \rho_X / \sqrt{\det(J_{g^{-1}, X}^T J_{g^{-1}, X})} \\
        = & \tau_{i,f} \sqrt{\frac{\det(J_{g, X}^T J_{g, X})}{\det(g_{g^{-1}, X}^T J_{g^{-1}, X})}}
\end{split}
\end{equation}
where $J_{g,X}$ is the Jacobian matrix with respect to $X$. In the last line of equation~(\ref{eq:backward_density}) the density $\rho_X$ has been substituted considering that $\mathbf{s}_f = g(\mathbf{s}_i; X)$. Furthermore, since $g^{-1}$ is the inverse of $g$ with respect to $\mathbf{s}_f$ one has:
\begin{equation*}
g^{-1}(g(\mathbf{s}_i; X); X) = \mathbf{s}_i
\end{equation*}
Taking the gradient with respect to $X$ one finds after a few manipulations:
\begin{equation*}
\det(J_{g^{-1}, \mathbf{s}_f})^2 = \frac{\det(J_{g^{-1},X}^T J_{g^{-1},X})}{\det(J_{g,X}^T J_{g,X})}
\end{equation*}
Substituting back into equation~(\ref{eq:backward_density}) one gets the very simple result that \mbox{$\rho_Y = \tau_{i,f} /\det(J_{g^{-1}, \mathbf{s}_f})$}, i.e. $Y$ is distributed as $\mathbf{s}_f$ up to a Jacobian factor $|\partial \mathbf{s}_i / \partial \mathbf{s}_f|$. With a Physicist's notation this writes as a conservation law, as $\tau_{i,f} d \mathbf{s}_f = \tau_{f, i} d \mathbf{s}_i$, with $\tau_{f,i} \equiv \rho_Y$. Thus:
\begin{equation*}
\begin{split}
\frac{1}{N} \sum_{k=1}^N{\omega_{i,k} \rho_i(\mathbf{s}_{i,k})} \simeq & \int{\rho_i(\mathbf{s}_i) \det(J_{g^{-1}, \mathbf{s}_f}) \rho_Y(\mathbf{s}_i; \mathbf{s}_f) d\mathbf{s}_i} \\
\simeq & \int{\tau_{i, f}(\mathbf{s}_f; \mathbf{s}_i) \rho_i(\mathbf{s}_i) d\mathbf{s}_i} = \rho_f(\mathbf{s}_f)
\end{split}
\end{equation*}
where in the first line the classical \ac{MC} result for weighted particles is used, with $\omega_{i,j} =  \det(J_{g^{-1}, \mathbf{s}_f})$, and in the second $\rho_Y$ has been substituted. This completes the proof.

\section{Proof of lemma~\ref{lm:transport_weight}}
\label{sec:proof_2}
Let us compute the backward weight from the forward process. It goes as the inverse of the Jacobian determinant of the transform $\mathbf{s}_i \rightarrow \mathbf{s}_f$, i.e.  $\omega_{i,f} = 1/\omega_{f,i} = 1/\det(J_{\mathbf{s}_f, \mathbf{s}_i})$. Let us write $\beta(\epsilon)$ the pre-generated \ac{CEL}. Then, the randomisation of the final kinetic energy, $\epsilon_f$, is given by equation~(\ref{eq:forward_MUM}). It does not depend on $\vec{r}_i$ nor on $\vec{u}_i$, therefore the forward Jacobian determinant can be factorised as $\omega_{f,i} = \omega_{i,f, \mathcal{E}} \omega_{i,f, G}$ with:
\begin{equation}  
\omega_{f,i, \mathcal{E}} = \left| \frac{\partial \epsilon_f}{\partial \epsilon_i} \right|\ \mathrm{and}\   
\omega_{f,i, G} = \left| \begin{array}{cc}
\frac{\partial \vec{r}_f}{\partial \vec{r}_i} & \frac{\partial \vec{r}_f}{\partial \vec{u}_i} \\
\frac{\partial \vec{u}_f}{\partial \vec{r}_i} & \frac{\partial \vec{u}_f}{\partial \vec{u}_i} 
\end{array} \right|
\end{equation}
The kinetic weight factor $\omega_{i,f,\mathcal{E}}$ was computed previously and is given by equation~(\ref{eq:interaction_backward_jacobian}). The second weight factor is geometric and depends {\it a priori} on the transport from $(\vec{r}_i, \vec{u}_i)$ to $(\vec{r}_{f}, \vec{u}_{f})$. However, it should not depend on the details of the stepping procedure, up to discretisation errors. The proof of lemma~\ref{lm:transport_weight} requires considering the continuous limit. Therefore, let us consider the following stepping procedure with $n$ transport steps, from $s_0=s_i$ to $s_n=s_f$, where the step length is given by a uniform column density increment, as:
\begin{equation}
\label{eq:uniform_step_length}
\frac{x(\epsilon_f; \epsilon_i)}{n} = \int_{s=d_j}^{d_{j+1}}{\rho(s) ds}
\end{equation}
with $x(\epsilon_f; \epsilon_i)$ the randomised column density between the two vertices. The stepping rule from state $\mathbf{s}_j$ to $\mathbf{s}_{j+1}$ can be written generically as:
\begin{equation}
\begin{split}
\vec{r}_{j+1} = & \vec{r}_j + \Delta d_j \vec{u}_j + \mathcal{O}(\frac{1}{n^2}) \\
\vec{u}_{j+1} = & \vec{u}_j + \vec{u}_j \times \vec{\Omega}_j + \mathcal{O}(\frac{1}{n^2}) 
\end{split}
\end{equation}
with $\Delta d_j = d_{j+1}-d_j$  and where $\vec{\Omega}_j$ generates an infinitesimal rotation, e.g. due to multiple scattering or an external magnetic field, depending on $\vec{r}_j$, $\vec{u}_j$ and $\epsilon_j$. The quantities $\Delta d_j$ and $\vec{\Omega}_j$ are both of order $\mathcal{O}(\frac{1}{n})$. At same order, the Jacobian for the step writes as a block matrix, as:
\begin{equation}
\begin{split}
M_{j+1,j} = & \left[ \begin{array}{cc}
I_3+\frac{\partial \Delta d_j}{\partial \vec{r}_j} \otimes \vec{u}_j & \nabla_u \otimes (\Delta d_j \vec{u}_j) \\
\nabla_r \otimes (\vec{u}_j \times \vec{\Omega}_j) & I_3 + \nabla_u \otimes (\vec{u}_j \times \vec{\Omega}_j) 
\end{array} \right] + \mathcal{O}(\frac{1}{n^2}) \\
 = & \left[ \begin{array}{cc}
A_j & B_j \\
C_j & D_j  
\end{array} \right]
\end{split}
\end{equation}
where $I_3$ is the identity matrix of rank 3 and with $\otimes$, the outer product. At order $\mathcal{O}(\frac{1}{n})$ the lower right block, $D_j$, is invertible and its inverse writes:
\begin{equation}
D_j^{-1} = I_3 - \nabla_u \otimes (\vec{u}_j \times \vec{\Omega}_j) + \mathcal{O}(\frac{1}{n^2}) 
\end{equation}
Following, the Schur complement of $M_{j,j+1}$ with respect to $D_j$ is simply given as:
\begin{equation}
M_{j+1,j}/D_j = I_3+\frac{\partial \Delta d_j}{\partial \vec{r}_j} \otimes \vec{u}_j + \mathcal{O}(\frac{1}{n^2})
\end{equation}
where we have neglected terms of order $\mathcal{O}(\frac{1}{n^2})$ or lower. The matrices $M_{j,j+1}$ and $D_j (M_{j,j+1}/D_j)$ have the same determinant. Therefore let us write:
\begin{equation}
\begin{split}
M_{f,i}^{(n)} = & \prod_{j=1}^n{\left( I_3 + \nabla_u \otimes (\vec{u}_j \times \vec{\Omega}_j) \right) \left( I_3 + \frac{\partial \Delta d_j}{\partial \vec{r}_j} \otimes \vec{u}_j \right)} \\
 = &  \prod_{j=1}^n{D_j \left( M_{j,j+1}/D_j \right)} + \mathcal{O}(\frac{1}{n})
\end{split}
\end{equation}
By construction, the matrices $M_{i,f}^{(n)}$ converge to a matrix whose determinant is the geometric weight factor $J_{i,f, G}$. Furthermore, according to the matrix determinant lemma, one has:
\begin{equation}
\begin{split}
\left| I_3 + \nabla_u \otimes (\vec{u}_j \times \vec{\Omega}_j) \right| = & 1 +\nabla_u \cdot (\vec{u}_j \times \vec{\Omega}_j) \\
 = & 1
\end{split}
\end{equation}
since $\nabla_u \cdot (\vec{u}_j \times \vec{\Omega}_j) = 0$. Similarly, one finds:
\begin{equation}
\begin{split}
\left|  I_3 + \frac{\partial \Delta d_j}{\partial \vec{r}_j} \otimes \vec{u}_j \right| = & 1 + \frac{\partial \Delta d_j}{\partial \vec{r}_j} \cdot \vec{u}_j \\
 = & \frac{\partial d_{j+1}}{\partial d_j} \\
 = & \frac{\rho(\vec{r}_{j+1})}{\rho(\vec{r}_j)}
\end{split}
\end{equation}
where the last lines falls from the expression of the step length given by equation~(\ref{eq:uniform_step_length}). Collecting results, one finally has:
\begin{equation}
\omega_{f,i,G} = \left| \lim_{n \rightarrow \infty }{M_{i,f}^{(n)}} \right|  = \frac{\rho(\vec{r}_i)}{\rho(\vec{r}_f)}
\end{equation}
which completes the proof.

\section{Proof of lemma~\ref{lm:flux_to_density}}
\label{sec:proof_3}
Let us first consider the initial state boundary case with no transverse transport, i.e. 1 dimensional propagation. Then, a \ac{MC} state can be summarised as $\mathbf{s} = \{\epsilon, s\}$ with $s$ the curvilinear abscissae along a straight line trajectory such that increasing values of $s$ correspond to a forward propagation. Let $s_I$ be the start position and $\phi_S(\epsilon_I)$ be the known initial flux. Although the initial state density is not specified before $s_I$ one can consider an equivalent case, with an initial state density $\rho_i(\epsilon_i, s_i)$ for $s_i \leq s_I$, and yielding the same transported spectrum, $\phi_S$, through the interface. For the sake of simplicity let us set the arbitrary medium density before the interface to $\rho(s_I)$. A procedure for generating the initial state density is as following. Let $\beta(\epsilon)$ be the pre-generated \ac{CEL} and $\epsilon(x)$ its generating function. Then, let us first draw the kinetic energy $\epsilon_I$ at the interface crossing according to $\phi_S(\epsilon_I)$ and read the corresponding column density, $x_I$, from $\epsilon(x_I) = \epsilon_I$. Secondly, the initial position is drawn from an {\it a priori} arbitrary distribution $\rho_d(s_i; \epsilon_I)$. Finally, the initial kinetic energy is set according to the generated \ac{CEL}, as $\epsilon_i = \epsilon(x_I-\rho_I (s_I-s_i))$. The corresponding initial state density goes as:
\begin{equation}
\label{eq:equivalent_density}
\begin{split}
\rho_i(\epsilon_i, s_i) = & \phi_S(\epsilon_I) \rho_d(s_i; \epsilon_I) \frac{\partial \epsilon_I}{\partial \epsilon_i} \\
 = & \phi_S(\epsilon_I) \rho_d(s_i; \epsilon_I) \frac{\beta(\epsilon_I)}{\beta(\epsilon_i)} 
\end{split}
\end{equation}
By construction this generation procedure would yield the \ac{MC} flux $\phi_S$ at the start interface if there were no interactions, only continuous energy loss. To be consistent, in a forward \ac{MC}, events that interact before the start interface can be discarded and their weight set to zero. Thus, in order to yield the correct flux at the interface the density given by equation~(\ref{eq:equivalent_density}) needs to be divided by the survival probability up to the interface, $P_{\lambda,\beta}(\epsilon_I; \epsilon_i)$, given in equation~(\ref{eq:survival_probability}). With this generation recipe the forward and backward procedures are identical at the interface crossing. Thus, a backward transported particle can be sampled before the initial interface according to lemma~\ref{lm:transport_weight} and using the equivalent initial state density, $\rho_i$.

Let us know consider a particular case where the initial positions $s_i$ are distributed according to the interaction probability density, as:
\begin{equation}
\rho_d(s_i; \epsilon_I) = \frac{\partial P_{\lambda,\beta}(\epsilon_I; \epsilon_i)}{\partial s_i} = P_{\lambda,\beta}(\epsilon_I; \epsilon_i) \frac{\rho(s_I)}{\lambda(\epsilon_i)}
\end{equation}
where one should recall that $\epsilon_i$ is actually a function of $s_i$. Following, the initial state density has a simple expression, as:
\begin{equation}
\begin{split}
\rho_i(\epsilon_i, s_i) = & \phi_S(\epsilon_I) \frac{\rho(s_I)}{\lambda(\epsilon_i)} \frac{\beta(\epsilon_I)}{\beta(\epsilon_i)} \\
 = & \phi_S(\epsilon_I) \frac{\rho(s_I)}{\lambda(\epsilon_I)} \frac{1}{\omega_{i, I}}
\end{split}
\end{equation}
where $\omega_{i,I}$ is the \ac{BMC} weight factor for the transport from the initial position to the interface, given by lemma~\ref{lm:transport_weight}. Thus, in a \ac{BMC} step a particle crossing the initial state interface samples the equivalent initial density, $\rho_i(\epsilon_i, s_i)$, with a weight given as:
\begin{equation}
\label{eq:equivalent_weight}
\rho_i(\epsilon_i, s_i) \omega_{i,f} = \phi_S(\epsilon_I) \frac{\rho(s_I)}{\lambda(\epsilon_I)} \omega_{I, f}
\end{equation}
Since this weight does not depend on the initial state properties $\{\epsilon_i, s_i\}$, it is useless to perform the backward transport before the interface. Everything goes as if the \ac{MC} particle would have originated from the interface with an equivalent initial state density as stated in lemma~\ref{lm:flux_to_density}.

With a similar reasoning one can prove the same result for the final state density. Given, a flux $\phi_S(\epsilon_F)$ through an interface at $s_F$, one can build an equivalent final state density, $\rho_f(\epsilon_f, s_f)$, such that:
\begin{equation}
\frac{\omega_{i,f}}{\rho_f(\epsilon_f, s_f)} = \frac{\lambda(\epsilon_F)}{\rho(s_F) \phi_S(\epsilon_F)} \omega_{i, F}
\end{equation}
Once again, everything goes as if the \ac{MC} particle would have interacted on the interface with an equivalent final state density as stated in lemma~\ref{lm:flux_to_density}.

The direct proof for arbitrary \ac{MC} trajectories, when there is transverse transport, is not obvious because building a realistic equivalent state density would {\it a priori} depend on the 3 dimensional topology now. However, an equivalent generation procedure does not necessarily require a realistic geometry. It only needs to yield the correct flux at the interface while conforming to the Physics of the transport. In order to simplify the discussion let us only consider the case of an initial state interface. For any initial state on the interface, $\mathbf{s}_I = \{\epsilon_I, \vec{r}_I, \vec{u}_I\}$, let us attach an initial state space of infinite extension connected at $\vec{r}_I$ with a uniform density $\rho(\vec{r}_I)$. The material is the same than the one after the interface, such that the Physics of the transport is unchanged. Since the initial state space is invariant by space translation and rotation  an initial state is summarised as $\{\epsilon_i, s_i\}$ where $s_i \leq s_I$ is the curvilinear abscissae along the \ac{MC} trajectory, whose details do not need to be simulated. Following, an initial state is generated as previously by drawing a state $\mathbf{s}_I = \{\epsilon_I, \vec{r}_I, \vec{u}_I\}$ on the interface and a curvilinear distance $d$ over $\rho_d(s_i;\epsilon_I)$ from which the initial kinetic energy, $\epsilon_i$ is computed, according to the precomputed continuous energy loss. The particle is transported in the initial state space until it reaches $s_I=s_i+d$ where it enters the true geometry at $\vec{r}_I$ with direction $\vec{u}_I$. The corresponding backward transport is straightforward. When the backward state reaches the interface it continues propagating in the same material with a uniform density and no boundaries. With this procedure and the same reasoning as for the previous unidimensional case, one obviously recovers the results of equation~(\ref{eq:equivalent_weight}).

\section{Proof of lemma~\ref{lm:vertex_weight}}
\label{sec:proof_4}
Let us consider the following two steps procedure for simulating the vertex transform. First the discrete energy loss, $q=\epsilon_i-\epsilon_f$ is randomised from the marginal macroscopic cross-section $\partial \sigma / \partial q$. Second, $\vec{u}_f$ is generated given $\vec{u}_i$, $\epsilon_i$ and $\epsilon_f$, e.g. according to the kinematics or from a conditional distribution. Following this procedure, the randomisation of $\epsilon_f$ does not depend on $\vec{u}_i$, since the differential cross-sections are invariant by rotation. Thus the backward weight can be factorised as:
\begin{equation}
\omega_{i,f} = \left| \frac{\partial \epsilon_i}{\partial \epsilon_f} \right| \left| \frac{\partial \vec{u}_i}{\partial \vec{u}_f} \right|
\end{equation}
The transform $\vec{u}_i \rightarrow \vec{u}_f$ is a pure rotation, $\mathcal{R}$. Again, since the differential cross-sections are invariant by rotation it does not depend on $\vec{u}_i$. It depends only on $\epsilon_i$, $\epsilon_f$ and some random numbers. Therefore its Jacobian determinant is $\det(\mathcal{R}) = 1$, which completes the proof.

\section{Proof of lemma~\ref{lm:multi_process}}
\label{sec:proof_5}
Let us write $\mathbf{s}^*_i$ the initial state at the interaction vertex but before the interaction occurs, $\mathbf{s}_i$ the state right after and $\mathbf{s}_f$ the final state at the next vertex, before interaction. Let us first prove that the proposed procedure yields the correct total interaction length $1/\lambda = 1/\lambda_1 + ... 1/\lambda_m$ while selecting the interacting process with the bias probability $p_\ell(\epsilon_i) = \lambda(\epsilon_i) / \lambda_\ell(\epsilon_i)$, instead of $p_\ell(\epsilon_i^*)$. An inductive procedure is used for the proof. Let us initialise the procedure by considering the case of two independent processes. Let us write $\epsilon_{i,1}$ and $\epsilon_{i, 2}$ the random variables with complementary \ac{cdf} $P_{\lambda_1, \beta}(\epsilon_i; \epsilon_f)$ and $P_{\lambda_2, \beta}(\epsilon_i; \epsilon_f)$. The corresponding \ac{pdf} write:
\begin{equation}
\begin{split}
\rho_{\lambda_\ell, \beta}(\epsilon_i; \epsilon_f) = & -\frac{\partial P_{\lambda_\ell, \beta}}{\partial \epsilon_i} \\
= & \frac{P_{\lambda_\ell, \beta}(\epsilon_i; \epsilon_f)}{\lambda_\ell(\epsilon_i) \beta(\epsilon_i)}
\end{split}
\end{equation}
Let us now consider the random variable $\mathcal{E}_i = \min(\mathcal{E}_{i,1}, \mathcal{E}_{i,2})$ and the plane $(\epsilon_{i,1}, \epsilon_{i, 2})$ for the joint occurence of $\mathcal{E}_{i,1}$ and $\mathcal{E}_{i,2}$. The event $\{ \mathcal{E}_i \geq \epsilon_i \}$ requires both $\{ \mathcal{E}_{i, 1} \geq \epsilon_i \}$ and $\{ \mathcal{E}_{i,2} \geq \epsilon_i \}$ which are independent events. Thus:
\begin{equation}
\begin{split}
P(\mathcal{E}_i \geq \epsilon_i | \epsilon_f) = & P(\mathcal{E}_{i,1} \geq \epsilon_i | \epsilon_f) P(\mathcal{E}_{i,2} \geq \epsilon_i | \epsilon_f) \\
= & P_{\lambda_1, \beta}(\epsilon_i; \epsilon_f) P_{\lambda_2, \beta}(\epsilon_i; \epsilon_f) \\
= & P_{\lambda_1 \oplus \lambda_2, \beta}(\epsilon_i; \epsilon_f)
\end{split}
\end{equation}
where $1/(\lambda_1 \oplus \lambda_2) = 1/\lambda_1 + 1/\lambda_2$. Therefore $\mathcal{E}_i$ has the same \ac{cdf} than a single process that would be randomised with the equivalent interaction length $\lambda = \lambda_1 \oplus \lambda_2$. Furthermore, for a given occurrence $\epsilon_i$ of $\mathcal{E}_i$ the probability $p_1(\epsilon_f;\epsilon_i)$ that the $1^\text{st}$ process was the minimiser of $\mathcal{E}_i$ writes:
\begin{equation}
\begin{split}
p_1(\epsilon_i; \epsilon_f) = & \frac{\rho_{\lambda_1, \beta}(\epsilon_i; \epsilon_f)}{\rho_{\lambda_1 \oplus \lambda_2, \beta}(\epsilon_i; \epsilon_f)} P(\mathcal{E}_{i,2} \geq \epsilon_i | \epsilon_f) \\
= & \frac{\rho_{\lambda_1, \beta}(\epsilon_i; \epsilon_f) P_{\lambda_2, \beta}(\epsilon_i; \epsilon_f)}{\rho_{\lambda_1 \oplus \lambda_2, \beta}(\epsilon_i; \epsilon_f)} \\
= & \frac{\lambda(\epsilon_i)}{\lambda_1(\epsilon_i)}
\end{split}
\end{equation}
which is indeed the expected bias probability.

Let us now consider an inductive step and assume that the randomisation procedure is correct for $m-1 \geq 2$ processes. Then, let us write $\mathcal{E}_i^{(m-1)} = \min(\mathcal{E}_{i,1}, ..., \mathcal{E}_{i, m-1})$. Following from the previous assumption, $\mathcal{E}_i^{(m-1)}$ is distributed as a single random variable with an equivalent interaction length $\lambda^{(m-1)} = \lambda_1 \oplus ... \oplus \lambda_{m-1}$ and with occurrence probability for the $\ell^\text{th}$ sub-process given as $p_\ell^{(m-1)}(\epsilon_i) = \lambda^{(m-1)}(\epsilon_i)/\lambda_\ell(\epsilon_i)$. In addition, since the procedure is in particular correct for $m=2$, the random variable $\mathcal{E}_i = \min(\mathcal{E}_i^{(m-1)}, \mathcal{E}_{i,m})$ is also distributed as a single random variable with equivalent interaction length $\lambda^{(m)} = \lambda^{(m-1)} \oplus \lambda_m$ and selection probabilities $p_{\ell<m}^{(m)} = \lambda^{(m)}/\lambda^{(m-1)} p_\ell^{(m-1)}$ or $p_m^{(m)}(\epsilon_i) = \lambda^{(m)}/\lambda^{(m-1)}$. Developing one finds $\lambda^{(m)} = \lambda$ and $p_\ell^{(m)}(\epsilon_i) = \lambda(\epsilon_i) / \lambda_\ell(\epsilon_i)$, which completes the inductive demonstration.

Finally, at each interaction vertex one needs to correct the backward weight from the bias factor due to the sub-process selection, as given by corollary~\ref{cor:backward_mixture_sampling}. The corresponding bias weight writes:
\begin{equation}
\omega_{b} = \frac{p_j(\epsilon_i^*)}{p_j(\epsilon_i)} \omega^{(j)}_{b} = \frac{\lambda(\epsilon_i^*) }{\lambda_j(\epsilon_i^*)} \frac{\lambda_j(\epsilon_i)}{\lambda(\epsilon_i)} \omega^{(j)}_{b}
\end{equation}
where $\omega^{(j)}_{b}$ is any additional bias due to the vertex transform of the sub-process. This is equivalent to substituting $\lambda$ by $\lambda_j$ and $\omega_b$ by $\omega_b^{(j)}$ in equation~(\ref{eq:vertex_weight}), which completes the proof.

\section{Proof of lemma~\ref{lm:adjoint_montecarlo}}
\label{sec:proof_6}
Let us consider the forward interaction process at the vertex and let us explicitly write $\tau_{i,f}$ the \ac{pdf} of the transform from initial state $\mathbf{s}_i$ to final state $\mathbf{s}_f$, as:
\begin{equation}
\tau_{i,f}(\epsilon_f; \epsilon_i) = \frac{1}{\sigma} \frac{\partial \sigma}{\partial \epsilon_f} \theta(\epsilon_f) \theta(\epsilon_i-\epsilon_f)
\end{equation}
where $\theta$ is the Heaviside step function. Let us write $\tau^\dagger$ the \ac{pdf} of the adjoint process, given as:
\begin{equation}
\begin{split}
\tau_{i,f}^\dagger(\epsilon_i; \epsilon_f) = & \frac{\tau(\epsilon_f; \epsilon_i) h_b(\epsilon_i, \epsilon_f) }{\int_{-\infty}^{+\infty}\tau(\epsilon_f; \epsilon_i') h_b(\epsilon_i', \epsilon_f)  d\epsilon_i'} \\
= & \tau(\epsilon_f; \epsilon_i) h_b(\epsilon_i, \epsilon_f) \frac{\lambda^\dagger(\epsilon_f)}{\lambda(\epsilon_i)}
\end{split}
\end{equation}
The corresponding \ac{cdf} is given by $P_{\sigma, h_b}$, defined in equation~(\ref{eq:adjoint_cdf}). It provides a possible randomisation process, $g^{\dagger, -1}$, for the initial kinetic energy, biased with respect to the direct inverse, $g^{-1}$ as defined in lemma~\ref{lm:backward_sampling}. With a similar reasoning than for the proof of lemma~\ref{lm:backward_sampling} one can show that the \ac{pdf} of the corresponding biased forward process, $g^\dagger$, writes:
\begin{equation}
\rho_{g^\dagger}(\epsilon_f; \epsilon_i) = \tau_{i,f}^\dagger(\epsilon_i; \epsilon_f) \left| \frac{\partial \epsilon_i}{\partial \epsilon_f} \right|
\end{equation}
Thus, the bias factor, $\omega_b$, in equation~(\ref{eq:vertex_weight}) goes as:
\begin{equation}
\omega_b(\epsilon_i; \epsilon_f) = \frac{\tau}{\rho_{g^\dagger}} = \frac{\lambda(\epsilon_i)}{\lambda^\dagger(\epsilon_f)} {\left| \frac{\partial \epsilon_f}{\partial \epsilon_i} \right|}\frac{1}{h_b(\epsilon_i, \epsilon_f)}
\end{equation}
Injecting this result in equation~(\ref{eq:vertex_weight}) yields equation~(\ref{eq:adjoint_weight}) which completes the proof.


\Urlmuskip=0mu plus 1mu\relax
\bibliographystyle{elsarticle-num}
\bibliography{biblio}

\end{document}